\documentclass[superscriptaddress,aps, prd, reprint, nofootinbib]{revtex4-2}

\usepackage{graphicx}% Include figure files
\usepackage{amsmath,amssymb}
\usepackage{dcolumn}% Align table columns on decimal point
\usepackage{bm}% bold math
\usepackage{xcolor}
\usepackage{soul}
\usepackage{tikz}

\usepackage{comment}

\usepackage[linktocpage=true]{hyperref}
\hypersetup{
colorlinks=true,
citecolor=darkblue,
linkcolor=reddish,
urlcolor=darkblue,
pdfauthor={},
pdftitle={},
pdfsubject={}
}

\newcommand{\ud}{\mathrm{d}}

\def\e{\varepsilon}

\definecolor{blue2}{cmyk}{1, 0.1, 0.1, 0.1}

\usepackage{colortbl}
\definecolor{lightgreen}{cmyk}{0.2, 0, 0.2, 0.2}
\definecolor{lightgray2}{cmyk}{0.1,0.1,0,0.1}
\definecolor{Red2}{RGB}{214, 39, 40}
\definecolor{Blue2}{RGB} {31, 119, 180}
\definecolor{Orange2}{RGB}{255, 127, 14}
\definecolor{Green2}{RGB}{44, 160, 44}
\definecolor{greyish2}{rgb}{.96,.96,.96}

\definecolor{Red}{RGB}{214, 39, 40}
\definecolor{Blue}{RGB} {31, 119, 180}
\definecolor{Orange}{RGB}{255, 153, 51}
\definecolor{Purple}{RGB}{178, 102, 255}
\definecolor{Green}{RGB}{44, 160, 44}
\definecolor{regal}{RGB}{90,0,120}
\definecolor{darkblue}{rgb}{0.15,0.35,0.55}
\definecolor{reddish}{rgb}{0.65, 0.2, 0.2}
\definecolor{darkgreen}{RGB}{50,150,0}
\definecolor{greyish}{rgb}{.90,.90,.90}
\definecolor{greyish2}{rgb}{.96,.96,.96}
\definecolor{greyish3}{rgb}{.37,.37,.37}
\definecolor{darkblue2}{rgb}{0.3,0.4,0.9}
\definecolor{Blue3}{RGB}{31, 119, 180}

\definecolor{pyBlue}{RGB}{31, 119, 180}
\definecolor{pyRed}{RGB}{214, 39, 40}
\definecolor{pyGreen}{RGB}{44, 160, 44}
\definecolor{pyBlue2}{RGB}{0, 111, 237}
\definecolor{pyRed2}{RGB}{224, 52, 36}
\definecolor{Mathematica1}{rgb}{0.368417, 0.506779, 0.709798}
\definecolor{Mathematica2}{rgb}{0.880722, 0.611041, 0.142051}

\def\beq{\begin{equation}}
\def\eeq{\end{equation}}

\usetikzlibrary{shapes.misc}

\tikzset{cross/.style={cross out, draw=black, minimum size=2*(#1-\pgflinewidth), inner sep=0pt, outer sep=0pt},
cross/.default={1pt}}

\begin{document}

%\preprint{APS/123-QED}

\title{Kinematic Flow and the Emergence of Time}

\author{Nima Arkani-Hamed}
\affiliation{Institute for Advanced Study, Princeton, NJ 08540, USA}

\author{Daniel Baumann}
\affiliation{Leung Center for Cosmology and Astroparticle Physics, Taipei 10617, Taiwan}
\affiliation{Center for Theoretical Physics, National Taiwan University, Taipei 10617, Taiwan}
\affiliation{Institute of Physics,
University of Amsterdam, Amsterdam, 1098 XH, The Netherlands}

\author{Aaron Hillman}
\affiliation{Walter Burke Institute for Theoretical Physics, Caltech, Pasadena, CA 91125, USA}
\affiliation{Department of Physics, Jadwin Hall, Princeton University, NJ 08540, USA}

\author{Austin Joyce}
\affiliation{Department of Astronomy and Astrophysics,
University of Chicago, Chicago, IL 60637, USA}
\affiliation{Kavli Institute for Cosmological Physics, 
University of Chicago, Chicago, IL 60637, USA}

\author{Hayden Lee}
\affiliation{Kavli Institute for Cosmological Physics, 
University of Chicago, Chicago, IL 60637, USA}

\author{Guilherme L.~Pimentel}
\affiliation{Scuola Normale Superiore and INFN, Piazza dei Cavalieri 7, 56126, Pisa, Italy}

\begin{abstract}
\noindent
Perhaps the most basic question we can ask about cosmological correlations is how their strength changes as we smoothly vary kinematic parameters. The answer is encoded in differential equations that govern this evolution in kinematic space. 
In this Letter, we introduce 
a new perspective on these differential equations.  We show that, in the simplified setting of conformally coupled scalars in a general FRW spacetime, the equations for arbitrary tree-level processes can be obtained from a small number of simple combinatorial rules. While this ``kinematic flow" is defined purely in terms of boundary data, it reflects the physics of bulk time evolution.  The unexpected regularity of the equations suggests the existence of an autonomously defined mathematical structure from which cosmological correlations, and the time evolution of the associated spacetime, emerge.

\end{abstract}

\maketitle

Cosmology gives us the strongest motivations to replace the seemingly foundational notion of spacetime with deeper, and likely more abstract, principles. From the breakdown of Einstein gravity coupled to matter near the Big Bang singularity, to the profound mysteries associated with the accelerated expansion of our Universe, it is cosmology that most powerfully calls for an overhaul of our description of the laws of Nature,  jettisoning the phrase ``time evolution" from the vocabulary of fundamental physics at the deepest level, and instead seeing the concept of time emerge as an extremely useful approximation when this becomes possible. 

\vskip2pt
It is also suggestive that our actual observations of the Universe are static, correlating measurements at different positions in the late Universe. We introduce a cosmological history to give a rational accounting of these spatial patterns, but ultimately the notion of ``cosmological time" is an auxiliary one, not present in the observables themselves. There should therefore be a description of the same physics that focusses only on the observed static correlations and in which time evolution becomes a derived concept.

\vskip2pt
Another motivation for reformulating cosmological perturbation theory in a way that dispenses with explicit evolution in time is that the standard  ``time-centered" computations 
are very complex.  This complexity can be traced to the time integrals over interactions in the quantum-mechanical sum over histories, and it likely conceals fascinating new physical and mathematical structures underlying cosmological correlators. 

\vskip2pt
For several years, these developments have been calling for the formulation of a radically new, ``timeless" approach to cosmological correlators. While some partial results in this direction have been found~\cite{Baumann:2022jpr,Maldacena:2011nz,Mata:2012bx,McFadden:2009fg,Bzowski:2013sza,Arkani-Hamed:2015bza,Arkani-Hamed:2018kmz,Baumann:2019oyu,Baumann:2020dch,Sleight:2019mgd,Sleight:2019hfp,Pajer:2020wxk,Goodhew:2020hob,Jazayeri:2021fvk,DiPietro:2021sjt,Hogervorst:2021uvp, Pimentel:2022fsc,Jazayeri:2022kjy,Wang:2022eop}---notably the ``cosmological polytopes" associated with a new combinatorial/geometric understanding of the flat-space wavefunction~\cite{Arkani-Hamed:2017fdk,Benincasa:2018ssx,Benincasa:2019vqr}---we have not yet seen a single example of a completely ``autonomous" set of ideas for determining the cosmological wavefunction, purely in terms of the kinematic data specifying the observable itself, capturing the dynamics of ``bulk'' time evolution in purely ``boundary'' terms.

\vskip2pt
Our goal in this Letter is to change this state of affairs. We will study an evolution in the space of kinematic configurations---called {\it kinematic flow}---that replaces the ordinary notion of cosmological time evolution. Remarkably, this flow can be defined autonomously in the space of boundary kinematics, and a few simple rules allow us to predict the differential equations for arbitrary tree graphs (at least in a toy model of cosmological evolution).

\vskip 4pt
\noindent
{\bf Boundary kinematics.}
We describe cosmological fluctuations
in Fourier space, where the kinematic data is a set of wavevectors (momenta)  $\vec{k}_a$, which form a closed polygon as a consequence of momentum conservation:
 \vspace{-6pt}
\begin{equation*}
\raisebox{-.5pt}{
 \begin{tikzpicture}[baseline=(current  bounding  box.center),scale=0.85]
 \draw[black, line width=1.pt] (3,-0.5) -- (3,0.75) -- (4.25,0.75) --  (4.25,-0.5) -- (3,-0.5);
    \node[left,color=black] at (3,0.125) {\footnotesize $\vec{k}_1$};
  \node[above,color=black] at (3.625,0.75) {\footnotesize $\vec{k}_2$};
    \node[right,color=black] at (4.25,0.125) {\footnotesize $\vec{k}_3$};
  \node[below,color=black] at (3.625,-0.5) {\footnotesize $\vec{k}_4$};
  \end{tikzpicture}
  }
  \hspace{1cm}
   \begin{tikzpicture}[baseline=(current  bounding  box.center),scale=.675]
\coordinate (1) at (0,1);
\coordinate (2) at (0.951,0.309);
\coordinate (3) at (0.588,-0.809);
\coordinate (4) at (-0.588,-0.809);
\coordinate (5) at (-0.951,0.309);
\draw[color=black, line width=1.pt] (1) -- (2) -- (3) -- (4) -- (5) -- (1); 
    \node[left,color=black] at (-.8,-0.25) {\footnotesize $\vec{k}_1$};
     \node[left,color=black] at (-.35,1.1) {\footnotesize $\vec{k}_2$};
       \node[right,color=black] at (.35,1.1) {\footnotesize $\vec{k}_3$};
   \node[right,color=black] at (.8,-0.25) {\footnotesize $\vec{k}_4$};
     \node[below,color=black] at (0,-0.8) {\footnotesize $\vec{k}_5$};
\end{tikzpicture}
 \vspace{-6pt}
\end{equation*}
Correlators are then functions of the lengths of the sides and diagonals of these kinematic polygons. 

\vskip 2pt
Certain kinematic configurations are special. First, we can analytically continue the side lengths, so that the perimeters of some sub-polygons (including the full polygon) vanish. As we will see, cosmological correlators have singularities at these locations~\cite{Maldacena:2011nz,Raju:2012zr,Arkani-Hamed:2017fdk}. It will be illuminating to represent these energy singularities graphically by shading the relevant sub-polygons, such as
\beq
\raisebox{-7pt}{\includegraphics[scale=0.7]{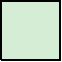}}
= k_{1234}\,,
\hspace{.6cm}
\raisebox{-7pt}{\includegraphics[scale=0.7]{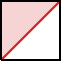}}= k_{12}+s\,,
\label{equ:Shading}
\eeq
where $k_{n_1\cdots n_j} \equiv |\vec{k}_{n_1}| +\cdots+ |\vec{k}_{n_j}|$ and $s \equiv |\vec{k}_1+\vec{k}_2|$.
Correlators can, in principle, also have singularities when the sum of some external edges become collinear with a diagonal.  We represent this by the following shading 
\beq
\raisebox{-7pt}{\includegraphics[scale=0.7]{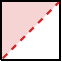}}= k_{12}-s\,,
\label{equ:Folded}
\eeq
where the dashed line indicates that the sign of the internal line is flipped when computing  the perimeter. 

\vskip4pt
\noindent
{\bf Graphs and tubings.} 
It is useful to introduce a bit of extra structure in kinematic space. Polygons beg to be triangulated, and so we will pick a particular triangulation of the kinematic polygon. 
 Any triangulation of a polygon has an associated dual graph that encodes the adjacency of triangles. For example, triangulations of the square correspond to graphs with two vertices
\begin{equation*}
 \begin{tikzpicture}[baseline=(current  bounding  box.center),scale=0.8]
\draw[thick,color=gray] (6, 0.125) -- (7.4, 0.125);
\draw[fill,color=Red] (6, 0.125) circle (.5mm);
\draw[fill,color=Blue] (7.4, 0.125) circle (.5mm);
   \draw[gray, line width=0.5pt] (3,-0.5) -- (4.25,0.75);
 \draw[Red, line width=1.pt] (3,-0.5) -- (3,0.75) -- (4.25,0.75);
  \draw[Blue, line width=1.pt] (3,-0.5) -- (4.25,-0.5) -- (4.25,0.75);
  \node[above,color=Red] at (6, 0.125)  {$X_1$};
  \node[above,color=Blue] at (7.4, 0.125)  {$X_2$};
  \node[below,color=gray] at (6.7, 0.125)  {$Y$};
\end{tikzpicture}
\end{equation*} 
where $X_1 \equiv k_{12}$, $X_2 \equiv k_{34}$ and $Y \equiv |\vec{k}_1+\vec{k}_2|$.
Higher-point triangulations have similar representations, such as
\begin{equation*}
\begin{tikzpicture}[line width=1.pt, scale=.7]
\coordinate (1) at (0,1);
\coordinate (2) at (0.951,0.309);
\coordinate (3) at (0.588,-0.809);
\coordinate (4) at (-0.588,-0.809);
\coordinate (5) at (-0.951,0.309);
\draw[gray, line width=0.5pt] (4) -- (1);
\draw[gray, line width=0.5pt] (4) -- (2);
\draw[color=Red, line width=1.pt] (4) -- (5) -- (1);
\draw[color=Orange, line width=1.pt] (1) -- (2); 
\draw[color=Blue, line width=1.pt] (4) -- (3) -- (2);
\draw[thick,color=gray] (2.4, 0.125) -- (5.2, 0.125);
\draw[fill,color=Red] (2.4, 0.125) circle (.5mm);
\draw[fill,color=Orange] (3.8, 0.125) circle (.5mm);
\draw[fill,color=Blue] (5.2, 0.125) circle (.5mm);
 \node[above,color=Red] at (2.4, 0.125)  {$X_1$};
 \node[above,color=Orange] at (3.8, 0.125)  {$X_2$};
  \node[above,color=Blue] at (5.2, 0.125)  {$X_3$};
 \node[above,color=gray] at (3.1, -0.55)  {$Y$};
  \node[above,color=gray] at (4.5, -0.55)  {$Y'$};
\end{tikzpicture}
\end{equation*}
where each vertex is associated to a triangle in the triangulation of the pentagon.

\vskip2pt
Above we described how the singularities of correlators are related to the perimeters of sub-polygons. This also has a useful representation on the dual graph. 
We first decorate the graph with {\it markings} (crosses on internal lines),  
and then draw connected {\it tubes} (circlings of the vertices and crosses) of these marked graphs.  
To each tubing, we then assign the sum of vertex energies enclosed by the tube and the energies of the internal lines piercing the tube. For tubes that intersect an internal line and enclose the corresponding cross, we flip the sign of the internal energy. For example, the singularities shown in (\ref{equ:Shading}) and (\ref{equ:Folded}) correspond to 
\begin{equation*}
 \raisebox{-1pt}{\includegraphics[scale=0.7]{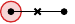}} \equiv\,   \raisebox{-7pt}{\includegraphics[scale=0.7]{Figures/Shadings/Square/sqpsisr}}
 \ , 
\hspace{0.4cm}
 \raisebox{-1pt}{\includegraphics[scale=0.7]{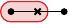}} 
 \equiv\,  \raisebox{-7pt}{\includegraphics[scale=0.7]{Figures/Shadings/Square/sqFsr}} \ ,
 \hspace{0.4cm}
  \raisebox{-1pt}{\includegraphics[scale=0.7]{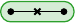}}  
  \,\equiv\, \raisebox{-7pt}{\includegraphics[scale=0.7]{Figures/Shadings/Square/sqZg}}\ .
\end{equation*}
These marked graphs and tubings will play an important role in the construction of differential equations.

\vskip4pt
\noindent
{\bf Differential equations.} 
The kinematic space of correlations can be expressed in terms of either a polygon or a marked graph. But these are completely static objects, how do they capture actual cosmological correlations? The most basic question we can ask is
how the strength of correlations changes
as we vary the external energies (i.e.~the shape of the polygon) infinitesimally. This is a differential question, and so we should expect that the answer is captured by a differential equation. What is more remarkable is that these differential equations can also be
 understood as arising from a sort of flow through the space of graph tubings. 

\vskip2pt
We denote the correlation function associated to a kinematic graph by $\psi$. We then take derivatives of $\psi$ with respect to the energies $Z_I$ (which includes both the external energies $X_i$ and the exchange energies $Y_j$).
Generically, this will generate new functions (and possibly the function $\psi$ itself). Differentiating these functions, produces additional new functions, and so on. A priori there is no reason that this procedure should ever terminate, but in our case it will. 
We therefore have a finite basis of functions $S_a$, and can assemble them into a vector $\vec{I} \equiv (\psi,S_1, \cdots ,S_{n-1})^T$. Since the system is finite, it must satisfy a differential equation of the form
\beq
\ud\hskip 1pt \vec{I} = A\, \vec{I}\, ,
\label{equ:DE}
\eeq
where $A$ is a matrix-valued one-form and we have defined the total differential $\ud \equiv \sum_I \partial_{Z_I} \ud Z_I $.

\vskip2pt
The challenge now is to generate the connection matrix $A$. In the following, we will describe a combinatorial procedure that accomplishes this task in terms of a {\it kinematic flow} through the space of (marked) graph tubings. The output of this procedure will be a connection of the  form
\beq
A = \sum_i \alpha_i\, \ud \log \Phi_i(Z)\,,
\label{equ:A-matrix}
\eeq 
where $\alpha_i$ are constant matrices and 
$\Phi_i(Z)$ are the {\it letters} of the differential equation. 
Notably, this equation is a sum of dlog forms and thus is Abelian flat ($\ud A =0$).

\vskip 2pt
Conceptually, there are three ingredients to the equation~\eqref{equ:DE}. First, we have to enumerate the possible letters that can appear in the connection matrix~\eqref{equ:A-matrix}. Next, we have to determine how many functions appear in the vector $\vec I$, and organize them in some way. Finally, we need a way to derive the differential of $\vec I$, and hence obtain the coefficient matrices~$\alpha_i$. We will address each of these needs in turn.

\vskip 4pt
\noindent
{\bf Letters.}  
We start by describing the possible singularities of the differential equations (i.e.~its letters). 
They are naturally encoded by the same graph tubings that we introduced to label kinematic limits. For example, the letters (represented as dlog forms) of the two-site chain are the following tubings of the marked graph:
\begin{equation*}
\begin{aligned}
&  \raisebox{-1pt}{\includegraphics[scale=0.7]{Figures/Tubings/two/letters/Lap}} \equiv  \ud \log(X_1+Y) \,,
\hspace{0.5cm}
 \raisebox{-1pt}{\includegraphics[scale=0.7]{Figures/Tubings/two/letters/Lam}} 
 \equiv  \ud \log(X_1-Y) \,,
\\
&\raisebox{-1pt}{\includegraphics[scale=0.7]{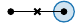}}   \equiv \ud \log(X_2+Y) \,,
\hspace{0.5cm}
 \raisebox{-1pt}{\includegraphics[scale=0.7]{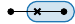}} 
  \equiv  \ud \log(X_2-Y)\,,
\\
& \raisebox{-1pt}{\includegraphics[scale=0.7]{Figures/Tubings/two/letters/Labg}}    \equiv  \ud \log(X_1+X_2)\,.
\end{aligned}
\label{equ:Letters-2Site}
\end{equation*}
Starting at three sites, vertices can connect to multiple lines, leading to 
letters with multiple sign flips, such as
 \begin{equation*}
 \begin{aligned}
    \raisebox{-1pt}{\includegraphics[scale=0.7]{Figures/Tubings/three/letters/LLbmp}}
     &\equiv \ud \log(X_2-Y+Y') \, , \\
          \raisebox{-1pt}{\includegraphics[scale=0.7]{Figures/Tubings/three/letters/LLbmm}}  &\equiv \ud \log(X_2-Y-Y')\, .
     \end{aligned}
    \end{equation*}
In total, there are 13 letters for the three-site graph.

\vskip 4pt
\noindent
{\bf Functions.}  Next, we define the set of basis functions of the differential equation. 
For this purpose, we introduce a second type of graph tubings.
This time, we draw all ``complete tubings" of the graph (non-nested and non-intersecting tubings that enclose all vertices). 
Each tubing corresponds to 
a unique basis function. Tubings that contain at least one cross are associated to source functions, while the wavefunction itself has no enclosed crosses.  
For the two-site chain, the set of functions is 
 \beq
 \begin{tikzpicture}[baseline=(current  bounding  box.center)]
  \node at (-0.65,0)  {$\psi$};
\node[inner sep=0pt] at (0,0)
   {\includegraphics[scale=0.7]{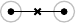}}; 
 \node at (1.3,0)  {$F$};
 \node[inner sep=0pt] at (2,0)
   {\includegraphics[scale=0.7]{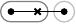}}; 
    \node at (1.3,-0.45)  {$\tilde F$};
    \node[inner sep=0pt] at (2,-0.5)
   {\includegraphics[scale=0.7]{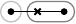}}; 
    \node at (3.2,0)  {$Z$};
 \node[inner sep=0pt] at (3.9,0)
   {\includegraphics[scale=0.7]{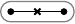}}; 
\end{tikzpicture}
\nonumber
\eeq
while, for the three-site chain, we have 
\beq
\begin{aligned}
\quad &\phantom{Q_1}\psi\,\   \raisebox{-1pt}{\includegraphics[scale=0.7]{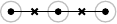}}  
&& \phantom{q_1} f\,\ \raisebox{-1pt}{\includegraphics[scale=0.7]{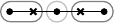}}  
 &&& \phantom{Z}g\,\ \raisebox{-1pt}{\includegraphics[scale=0.7]{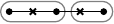}} 
\\
&&& \phantom{f} q_1\,\ \raisebox{-1pt}{\includegraphics[scale=0.7]{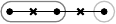}} 
&&& \phantom{Z}\tilde g\,\ \raisebox{-1pt}{\includegraphics[scale=0.7]{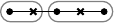}} 
 \\
\quad & \phantom{Q_1}F\,\   \raisebox{-1pt}{\includegraphics[scale=0.7]{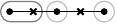}}   
\quad && \phantom{f}q_2\,\ \raisebox{-1pt}{\includegraphics[scale=0.7]{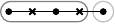}} 
&&&\phantom{g}Z\,\ \raisebox{-1pt}{\includegraphics[scale=0.7]{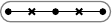}} 
\\
& \phantom{Q_1}\tilde F\,\  \raisebox{-1pt}{\includegraphics[scale=0.7]{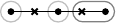}}  
&& \phantom{f}q_3\,\ \raisebox{-1pt}{\includegraphics[scale=0.7]{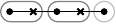}} 
\\
& \phantom{F} Q_{1}\,\ \raisebox{-1pt}{\includegraphics[scale=0.7]{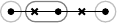}}  
&& \phantom{f}\tilde q_1\,\  \raisebox{-1pt}{\includegraphics[scale=0.7]{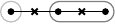}} 
\\
& \phantom{F} Q_{2} \,\ \raisebox{-1pt}{\includegraphics[scale=0.7]{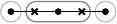}}  && \phantom{f}\tilde q_2\,\ \raisebox{-1pt}{\includegraphics[scale=0.7]{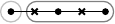}}  
\\
& \phantom{F} Q_3\,\ \raisebox{-1pt}{\includegraphics[scale=0.7]{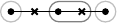}}  
&& \phantom{f}\tilde q_3\,\ \raisebox{-1pt}{\includegraphics[scale=0.7]{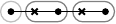}} 
\label{equ:SourceFunctions}
\end{aligned}
\nonumber
\eeq
Here, we have ordered the graphs by the number of vertices enclosed by tubes that also contain a cross.

\vskip 4pt
\noindent
{\bf Kinematic flow.} The final ingredients are the coefficient matrices $\alpha_i$ in \eqref{equ:A-matrix}. In the following, we introduce a set of {\it rules} that dictate how the graph tubings associated to the basis functions ``evolve" when we take derivatives. This will allow us to write down---by hand---the differential equations for all tree graphs.

\vskip2pt
We start with the tubing associated to a ``parent function" of interest and then generate a ``family tree" of its ``descendants" according to a set of  
rules, from which we can then read off the differential of the parent function.
The procedure is best illustrated by an example, and a particularly instructive case is the function $Q_1$ of the three-site chain:\\[4pt]
1.\,{\it Activation}---We first move through the graph and ``activate” each tube enclosing a vertex. Each activation forms a branch of the tree.

\vskip 2pt
For the function $Q_1$, we generate
 \beq
 \raisebox{-16pt}{\begin{tikzpicture}[scale=0.7]
\node[inner sep=0pt] at (0.6,0)
    {\includegraphics[scale=0.7]{Figures/Tubings/three/tree/threeQ1}}; 
\draw [color=gray,-stealth] (1.8,0) -- (2.6,0);
\draw [color=gray,-stealth] (1.8,0.15) -- (2.6,0.6);
\draw [color=gray,-stealth] (1.8,-0.15) -- (2.6,-0.6);
\node[inner sep=0pt] at (3.8,0.75)
    {\includegraphics[scale=0.7]{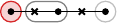}};
\node[inner sep=0pt] at (3.8,0)
    {\includegraphics[scale=0.7]{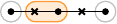}}; 
\node[inner sep=0pt] at (3.8,-0.75)
    {\includegraphics[scale=0.7]{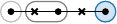}};   
\end{tikzpicture}}
\label{equ:Q1}
\eeq
where the colored tubes are activated in each branch.

\vskip 4pt
\noindent
2.\,{\it Growth}---Activated tubes with no crosses can “grow” to enclose adjacent crosses in all possible ways.
For example, in the bottom branch of~\eqref{equ:Q1}, we have
 \beq
 \begin{tikzpicture}[scale=0.7]
\node[inner sep=0pt] at (3.8,-0.5)
    {\includegraphics[scale=0.7]{Figures/Tubings/three/tree/threeQ1c}};   
\draw [color=gray,-stealth] (5,-0.5) -- (5.8,-0.5);
\node[inner sep=0pt] at (7.,-0.5)
    {\includegraphics[scale=0.7]{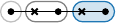}};
\end{tikzpicture}
\eeq

%\vskip 2pt
\noindent
3.\,{\it Merger}---If a grown tube intersects another tube, they “merge” (i.e.~their union becomes activated).
Such a merger occurs in the top branch of (\ref{equ:Q1}):
 \beq
 \begin{tikzpicture}[scale=0.7]
\node[inner sep=0pt] at (3.8,0.5)
    {\includegraphics[scale=0.7]{Figures/Tubings/three/tree/threeQ1a}};
\draw [color=gray,-stealth] (5.,0.5) -- (5.8,0.5);
\node[inner sep=0pt] at (7.,0.5)
    {\includegraphics[scale=0.7]{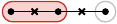}};
\end{tikzpicture}
\eeq
\vskip -4pt
\noindent
where {\raisebox{0pt}{\includegraphics[scale=0.7]{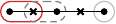}} {\raisebox{1.5pt}{$\Rightarrow$} \raisebox{0pt}{\includegraphics[scale=0.7]{Figures/Tubings/three/tree/threeq1ab}}\,.}  Tubes that grow/merge lead to additional descendant layers in the tree.

\vskip 4pt
\noindent
 4.\,{\it Absorption}---If an activated tube is adjacent to another tube with a cross, it further merges with this tube (``absorbs" it). This absorption is directional and only occurs if the cross in the activated tube points in the direction of the other tube. Absorption again generates an additional level of the tree.
 
 \vskip 2pt
 For the function $Q_1$, such an absorption happens once:
 \beq
 \begin{tikzpicture}[scale=0.7]
\node[inner sep=0pt] at (7.,-0.75)
    {\includegraphics[scale=0.7]{Figures/Tubings/three/tree/threeqt3c}};
\draw [color=gray,-stealth] (8.2,-0.75) -- (9.,-0.75);
\node[inner sep=0pt] at (10.2,-0.75)
    {\includegraphics[scale=0.7]{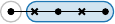}};
\end{tikzpicture}
\eeq
(Note there is no absorption for  {\raisebox{0pt}{\includegraphics[scale=0.7]{Figures/Tubings/three/tree/threeQ1b}} in \eqref{equ:Q1}, because the adjacent non-active tube does not contain a cross.)

\vskip 2pt
Combining the above elements,
the
complete family tree for the function $Q_1$ is 
 \beq
 \begin{tikzpicture}[scale=0.7]
\node[inner sep=0pt] at (0.6,0)
    {\includegraphics[scale=0.7]{Figures/Tubings/three/tree/threeQ1}}; 
\draw [color=gray,-stealth] (1.8,0) -- (2.6,0);
\draw [color=gray,-stealth] (1.8,0.15) -- (2.6,0.6);
\draw [color=gray,-stealth] (1.8,-0.15) -- (2.6,-0.6);
\node[inner sep=0pt] at (3.8,0.75)
    {\includegraphics[scale=0.7]{Figures/Tubings/three/tree/threeQ1a}};
\node[inner sep=0pt] at (3.8,0)
    {\includegraphics[scale=0.7]{Figures/Tubings/three/tree/threeQ1b}}; 
\node[inner sep=0pt] at (3.8,-0.75)
    {\includegraphics[scale=0.7]{Figures/Tubings/three/tree/threeQ1c}}; 
\draw [color=gray,thick,line width=0.5pt, dashed] (2.75,-1.03) -- (2.75,1.02) --  (4.85,1.02) -- (4.85,-1.03) -- (2.75,-1.03);  
\node[below] at (3.8,-1.03)  {$Q_1$};   
\draw [color=gray,-stealth] (5.,0.75) -- (5.8,0.75);
\draw [color=gray,-stealth] (5,-0.75) -- (5.8,-0.75);
\node[inner sep=0pt] at (7.,0.75)
    {\includegraphics[scale=0.7]{Figures/Tubings/three/tree/threeq1ab}};
            \node[above] at (7,1.02)  {$q_1$}; 
\node[inner sep=0pt] at (7.,-0.75)
    {\includegraphics[scale=0.7]{Figures/Tubings/three/tree/threeqt3c}};
        \node[below] at (7,-1.03)  {$\tilde q_3$}; 
\draw [color=gray,-stealth] (8.2,-0.75) -- (9.,-0.75);
\node[inner sep=0pt] at (10.2,-0.75)
    {\includegraphics[scale=0.7]{Figures/Tubings/three/tree/threeqt2bc}};
    \node[below] at (10.2,-1.03)  {$-\tilde q_2$}; 
\end{tikzpicture}
\nonumber
\eeq
To each graph in the tree, we assigned the function corresponding to the graph tubing (ignoring whether tubes are active or not). The sign of each descendant function is determined by the number of absorptions along the path from the original parent graph. We pick up a minus sign for each absorption. (This is why the function $\tilde q_2$ above comes with a minus sign.) If an activated tube at the first level contains more than one vertex, we multiply the corresponding function by the number of enclosed vertices (see Appendix~\ref{app:3site} for an example).

\vskip 2pt
Given a family tree like the one above, it is then 
simple to write down the differential of the parent function:
\vspace{-3pt}
\begin{enumerate}
\item For each graph in the tree, write the $\ud\log$ of the letter associated to the active tube.\vspace{-3pt}

\item Multiply this letter by the function associated to the graph {\it minus} the functions associated to all of its immediate descendant graphs, with an overall constant factor $\e$.
\end{enumerate}
\vspace{-3pt}
 \noindent
 For the function $Q_1$, this procedure gives
\beq
\begin{aligned}
\ud Q_1
&=     \e\, \big[
(Q_1 -q_1)
    \,\raisebox{-1pt}{\includegraphics[scale=0.7]{Figures/Tubings/three/letters/LLap}} 
+ q_1\,\raisebox{-1pt}{\includegraphics[scale=0.7]{Figures/Tubings/three/letters/LLabpr}} 
 \\ 
&\hspace{1.5cm} +    
 Q_{1} \raisebox{-1pt}{\includegraphics[scale=0.7]{Figures/Tubings/three/letters/LLbmp}} \\
&\hspace{0.5cm}+(Q_1 - \tilde q_{3})   \raisebox{-1pt}{\includegraphics[scale=0.7]{Figures/Tubings/three/letters/LLcp}}   +  (\tilde q_3  + \tilde q_2)
  \raisebox{-1pt}{\includegraphics[scale=0.7]{Figures/Tubings/three/letters/LLcm}}  \\
&\hspace{4.77cm}   -   \tilde q_2
  \raisebox{-1pt}{\includegraphics[scale=0.7]{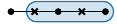}} \, \big].
\end{aligned}
\label{equ:dQ1}
\eeq
The overall factor of $\e$ normalizes the residues of the singularities. Later, we will see that this is related to 
the evolution of
the scale factor in a power-law cosmology.

\vskip 2pt
We have presented the simplest example which still displays all the important qualitative features. 
Of course, in more complicated graphs, repeated application of these steps can lead to more complex phenomena. For example, if an activated tube is connected to multiple internal lines it can can grow in multiple directions (see Appendix~\ref{app:3site}). 
Similarly, multiple absorptions can occur either at the same time or sequentially~\cite{Companion}.

\vskip 2pt
A critical feature of the kinematic flow rules is that they imply $\ud^2=0$ acting on any function, so that the equations are compatible. This requires a nontrivial interplay between the procedure itself (the subtraction of functions and assignment of signs) and linear relations between the kinematic letters that imply that certain combinations of $\ud\log$ forms vanish~\cite{Companion}.

\vskip 4pt
\noindent
{\bf Examples.} Remarkably, the preceding algorithm can be used to generate the system of differential equations associated to arbitrary tree graphs. For example, the system satisfied by the two-site chain is
\begin{align}
\label{eq:twositedpsi}
\begin{split}
\ud \psi &= \e\, \big[ (\psi - F)  \,\raisebox{-1pt}{\includegraphics[scale=0.7]{Figures/Tubings/two/letters/Lap}} \,+\, (\psi-\tilde F)
 \raisebox{-1pt}{\includegraphics[scale=0.7]{Figures/Tubings/two/letters/Lbp}} \\
&\hspace{1.32cm} + F \,  \raisebox{-1pt}{\includegraphics[scale=0.7]{Figures/Tubings/two/letters/Lam}}
\hspace{1.08cm} + \tilde F 
 \raisebox{-1pt}{\includegraphics[scale=0.7]{Figures/Tubings/two/letters/Lbm}} \, \big]
\end{split} \\[6pt]
%%%%%%%%%%%%%%%%%%%%%%%%%%%%%
\ud F 
& =    \e\, \big[ F 
  \,\raisebox{-1pt}{\includegraphics[scale=0.7]{Figures/Tubings/two/letters/Lam}}
 \, +  \,  (F-Z)\, \raisebox{-1pt}{\includegraphics[scale=0.7]{Figures/Tubings/two/letters/Lbp}}
\, + \, 
Z \,  \raisebox{-1pt}{\includegraphics[scale=0.7]{Figures/Tubings/two/letters/Labg}}\, \big] \label{equ:TwoSite-dF}
  \\[4pt]
 %%%%%%%%%%%%%%%%%%%%%%%%%%%%%
\ud \tilde F 
&=  \e\, \big[ \tilde F 
 \,\raisebox{-1pt}{\includegraphics[scale=0.7]{Figures/Tubings/two/letters/Lbm}}
 \, +  \,  (\tilde F-Z)  
\, \raisebox{-1pt}{\includegraphics[scale=0.7]{Figures/Tubings/two/letters/Lap}}
 \, + \, 
Z   \, \raisebox{-1pt}{\includegraphics[scale=0.7]{Figures/Tubings/two/letters/Labg}}\, \big]
  \\[6pt]
\ud Z 
& =    2\e\,Z  \, \raisebox{-1pt}{\includegraphics[scale=0.7]{Figures/Tubings/two/letters/Labg}}
\label{equ:TwoSite-dZ}
\end{align}
We can see the features of the flow through kinematic space in these equations. The differential of $\psi$ generates source functions $F,\tilde F$, which themselves are sourced by~$Z$, and which are associated to the growth of a tubing on the graph associated to the kinematic variables. 

\vskip 2pt
In Appendix~\ref{app:3site}, we present the complete set of equations for the three-site chain. Examples of longer chains and more complicated graph topologies 
can be found in~\cite{Companion}. The crucial point is that the equations for larger graphs  become very complicated, but the rules for generating these equations remain simple.

\vskip 4pt
\noindent
{\bf Emergent time.}  So far, the function $\psi$ has been a purely mathematical object, and we have not yet connected it to physics.  
Remarkably, $\psi$ has an interpretation as the wavefunction coefficient for a conformally coupled scalar in an FRW cosmology, with the scale factor $a(\eta) \propto \eta^{-(1+\varepsilon)}$~\cite{Arkani-Hamed:2017fdk}. One way to see this is to recast the output of a bulk calculation in a form where the system of equations like~\eqref{eq:twositedpsi}--\eqref{equ:TwoSite-dZ} can be derived and all of the basis functions can be given a bulk interpretation~\cite{Companion}. However, showing this explicitly would take us too far afield, so instead we will just give a glimpse of the encoding of time evolution by the differential equations.

\vskip2pt
From the general structure of the equations~\eqref{eq:twositedpsi}--\eqref{equ:TwoSite-dZ}, it is guaranteed that $\psi$ satisfies a fourth-order homogeneous differential equation (since the basis is four-dimensional and each function satisfies a first-order equation). However, in   the (non-generic) physical situation, two small miracles occur. The wavefunction $\psi$ actually satisfies a second-order inhomogeneous equation (or a third-order homogeneous equation):
\beq
\square\,
\psi= g \,(X_1+X_2)^{-(1-2\e)}\, ,
\eeq
and the relevant differential operator involves {\it only} $X_1$: 
\beq
\square \equiv (X_1^2-Y^2)\partial^2_{X_1}+2(1-\e) X_1\partial_{X_1}-\e(1-\e)\,,
\eeq
(and similarly for $X_2$). The nontrivial feature that this equation encodes is that there is a {\it local} differential operator that simplifies the singularity structure of the wavefunction $\psi$. This reflects the fact that the wavefunction can equally be thought of as arising from a bulk time integral associated to the Feynman diagram
\beq
\scalebox{1.0}{
 \raisebox{-26pt}{
\begin{tikzpicture}[line width=1. pt, scale=1.75]
\draw[fill=black] (0,0) -- (1,0);
\draw[fill=black] (0,0) -- (1,0);
\draw[line width=1.pt,lightgray] (0,0) -- (-0.25,0.55);
\draw[line width=1.pt,lightgray] (0,0) -- (0.25,0.55);
\draw[line width=1.pt,lightgray] (1,0) -- (0.75,0.55);
\draw[line width=1.pt,lightgray] (1,0) -- (1.25,0.55);
\draw[lightgray, line width=2.pt] (-0.5,0.55) -- (1.5,0.55);
\draw[fill=Red,Red] (0,0) circle (.03cm);
\draw[fill=Blue,Blue] (1,0) circle (.03cm);
\node[scale=1] at (0,-.17) {$X_1$};
\node[scale=1] at (1,-.17) {$X_2$};
\end{tikzpicture}
} 
}
\nonumber
\eeq
where the existence of such a differential operator is a consequence of the fact that the 
bulk-to-bulk propagator satisfies $\square G(x) = i\delta(x)$~\cite{Arkani-Hamed:2015bza}, where $\square$ is now the d'Alembertian in the bulk spacetime. It is remarkable that by asking a completely static question in the space of kinematic variables, we are led to a function that can equally well be interpreted as coming from cosmological time evolution.
}

\vskip 4pt
\noindent
{\bf Sum over graphs.} So far, we have discussed the differential equations for individual graphs (or specific triangulations of the kinematic polygon). For scattering amplitudes, however, the most stunning simplifications arise in the sum over Feynman graphs~\cite{Parke:1986gb,Benincasa:2007xk,Hodges:2012ym,Cheung:2014dqa,Arkani-Hamed:2017mur}, while the individual graphs are often not even physically meaningful objects. It remains an important open problem to find similar irreducible structures for the cosmological wavefunction that emerge when summing over graphs.

\vskip2pt
We will see some hints of deeper structure by removing the scaffolding of a choice of triangulation of the kinematic polygon, and instead summing over triangulations. 
The output will again have a physical interpretation, but now as the (flavor-ordered) wavefunction in ${\rm tr}\,\phi^3$ theory, which involves a sum over planar graphs~\cite{Cachazo:2013iea,Arkani-Hamed:2023lbd}. 

\vskip 2pt
A four-point function 
has two different triangulations
\begin{equation*}
 \begin{tikzpicture}[baseline=(current  bounding  box.center),scale=0.6]
   \draw[gray, line width=0.5pt] (3,-0.5) -- (4.25,0.75);
 \draw[black, line width=1.pt] (3,-0.5) -- (3,0.75) -- (4.25,0.75);
  \draw[black, line width=1.pt] (3,-0.5) -- (4.25,-0.5) -- (4.25,0.75);
\begin{scope}[xshift=3cm]
   \draw[gray, line width=0.5pt] (3,0.75) -- (4.25,-0.5);
 \draw[black, line width=1.pt] (4.25,-0.5) -- (3,-0.5) -- (3,0.75);
  \draw[black, line width=1.pt]  (3,0.75) -- (4.25,0.75) -- (4.25,-0.5);
\end{scope}
\end{tikzpicture}
\end{equation*}
corresponding to $s$ and $t$-channel exchanges.
Similarly, a five-point function is associated to a pentagon, which has five different triangulations
\beq 
\begin{tikzpicture}[line width=0.75pt, scale=.5]
\coordinate (1) at (0,1);
\coordinate (2) at (0.951,0.309);
\coordinate (3) at (0.588,-0.809);
\coordinate (4) at (-0.588,-0.809);
\coordinate (5) at (-0.951,0.309);
\draw[gray, line width=0.5pt] (4) -- (1);
\draw[gray, line width=0.5pt] (4) -- (2);
\draw[black, line width=1.pt] (1) -- (2) -- (3) -- (4) -- (5) -- (1);
\end{tikzpicture}
\hspace{15pt}
\begin{tikzpicture}[line width=0.75pt, scale=.5]
\coordinate (1) at (0,1);
\coordinate (2) at (0.951,0.309);
\coordinate (3) at (0.588,-0.809);
\coordinate (4) at (-0.588,-0.809);
\coordinate (5) at (-0.951,0.309);
\draw[gray, line width=0.5pt] (5) -- (2);
\draw[gray, line width=0.5pt] (5) -- (3);
\draw[black, line width=1.pt] (1) -- (2) -- (3) -- (4) -- (5) -- (1);
\end{tikzpicture}
\hspace{15pt}
\begin{tikzpicture}[line width=0.75pt, scale=.5]
\coordinate (1) at (0,1);
\coordinate (2) at (0.951,0.309);
\coordinate (3) at (0.588,-0.809);
\coordinate (4) at (-0.588,-0.809);
\coordinate (5) at (-0.951,0.309);
\draw[gray, line width=0.5pt] (4) -- (1);
\draw[gray, line width=0.5pt] (3) -- (1);
\draw[black, line width=1.pt] (1) -- (2) -- (3) -- (4) -- (5) -- (1);
\end{tikzpicture}
\hspace{15pt}
\begin{tikzpicture}[line width=0.75pt, scale=.5]
\coordinate (1) at (0,1);
\coordinate (2) at (0.951,0.309);
\coordinate (3) at (0.588,-0.809);
\coordinate (4) at (-0.588,-0.809);
\coordinate (5) at (-0.951,0.309);
\draw[gray, line width=0.5pt] (5) -- (2);
\draw[gray, line width=0.5pt] (4) -- (2);
\draw[black, line width=1.pt] (1) -- (2) -- (3) -- (4) -- (5) -- (1);
\end{tikzpicture}
\hspace{15pt}
\begin{tikzpicture}[line width=0.75pt, scale=.5]
\coordinate (1) at (0,1);
\coordinate (2) at (0.951,0.309);
\coordinate (3) at (0.588,-0.809);
\coordinate (4) at (-0.588,-0.809);
\coordinate (5) at (-0.951,0.309);
\draw[gray, line width=0.5pt] (3) -- (1);
\draw[gray, line width=0.5pt] (3) -- (5);
\draw[black, line width=1.pt] (1) -- (2) -- (3) -- (4) -- (5) -- (1);
\end{tikzpicture}
\nonumber
\eeq
In general, the triangulations of an $n$-gon are counted by the Catalan numbers $C_{n-2}$. 

\vskip2pt
The full wavefunction is again a member of a finite space of functions,
 which satisfy differential equations. 
 This time, we take derivatives with respect to the external energies $k_i$ (i.e.~the side lengths of the kinematic polygon).
The letters in the differential equation are now associated to shaded sub-polygons, where we allow internal lines to be either solid or dashed (see above).  Similarly, functions are associated to (possibly disconnected) shadings of polyangulations, which are the analogue of the disjoint tubings in the single-graph case.

 \vskip 2pt
 The kinematic flow now determines the evolution of shaded polyangulations.
 The precise rules 
 are given in~\cite{Companion}. Here, we will simplify the discussion by restricting ourselves to taking successive partial derivatives with respect to a single edge of the polygon (say the edge corresponding to the energy $k_1$).
The $k_1$-derivative of the wavefunction $\psi$ then generates shadings of all triangles that contain the edge associated to $k_1$. Internal lines can be solid or dashed.
 Taking further $k_1$-derivatives, the shadings grow by attaching further triangles.
 For example, in the case of the four-point function, the evolution of shadings is
 \beq
 \begin{tikzpicture}[baseline=-3pt,scale=.95]
\node[inner sep=0pt] at (0,0)
    {\includegraphics[scale=0.6]{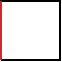}};
\draw [color=Red,thick,-stealth] (0.6,0.2) -- (1.2,0.65);
\draw [color=Red,thick,-stealth] (0.6,-0.2) -- (1.2,-0.65);
   \node[above] at (.7,.45) {$\textcolor{black}{\partial_{k_1}}$};
    \node at (-.6,0) {$\psi$};
	 \node[above] at (1.8,1.) {$\psi^{(s)}$};
\node[below] at (1.8,-1.) {$\psi^{(t)}$};
	 \node[above] at (3.6,1.) {$F^{(s)}$};
\node[below] at (3.6,-1.) {$F^{(t)}$};
    \node at (6,0) {$Z$};
\node[inner sep=0pt] at (1.8,0.7)
    {\includegraphics[scale=0.6]{Figures/Shadings/Square/sqpsisr}};
\node[inner sep=0pt] at (1.8,-0.7)
    {\includegraphics[scale=0.6]{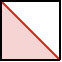}};
   \draw [color=gray,thick,-stealth] (2.4,0.7) -- (3.0,0.7);
   \draw [color=gray,thick,-stealth] (2.4,-0.7) -- (3.0,-0.7);
\node[inner sep=0pt] at (3.6,0.7)
    {\includegraphics[scale=0.6]{Figures/Shadings/Square/sqFsr}};
\node[inner sep=0pt] at (3.6,-0.7)
    {\includegraphics[scale=0.6]{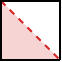}};
\draw [color=Red,thick,-stealth] (4.2,0.65) -- (4.8,0.2);
\draw [color=Red,thick,-stealth] (4.2,-0.65) -- (4.8,-0.2);
\node[inner sep=0pt] at (5.4,0)
    {\includegraphics[scale=0.6]{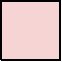}};
\end{tikzpicture}
\label{equ:4pt-Shared}
\eeq
There are two different paths for reaching the completely shaded square.
The function $Z$ associated to this shading is then ``shared" between the $s$- and $t$-channel equations.

\vskip2pt
For the four-point function, only one function is shared between channels, but the structure becomes more intricate at higher points~\cite{Companion}. 
For example, the five-point function has the following sequence of shadings: 
%
 %\vspace{-10pt}
 \beq
 \begin{tikzpicture}[baseline=-2pt,scale=.95]
\node[inner sep=0pt] at (0,0)
    {\includegraphics[scale=0.5]{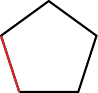}};
\draw [color=regal,thick,-stealth] (0.6,0.2) -- (1.2,0.6);
\draw [color=regal,thick,-stealth] (0.6,0) -- (1.2,0);
\draw [color=regal,thick,-stealth] (0.6,-0.2) -- (1.2,-.6);
\node[inner sep=0pt] at (1.8,.9)
    {\includegraphics[scale=0.5]{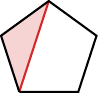}};
\node[inner sep=0pt] at (1.8,0)
    {\includegraphics[scale=0.5]{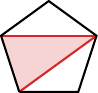}};
\node[inner sep=0pt] at (1.8,-.9)
    {\includegraphics[scale=0.5]{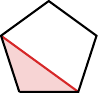}};
\draw [color=Green,thick,-stealth] (2.3,1.05) -- (3.1,1.6);
\draw [color=Green,thick,-stealth] (2.3,0.75) -- (3.1,.2);
\draw [color=Orange,thick,-stealth] (2.3,-0.75) -- (3.1,-.2);
\draw [color=Orange,thick,-stealth] (2.3,-1.05) -- (3.1,-1.6);
\draw [color=Red,thick,-stealth] (2.3,.15) -- (3.15,1.3);
\draw [color=Red,thick,-stealth] (2.3,-.15) -- (3.15,-1.3);
\node[inner sep=0pt] at (3.6,1.7)
    {\includegraphics[scale=0.5]{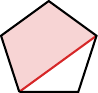}};
\node[inner sep=0pt] at (3.6,0)
    {\includegraphics[scale=0.5]{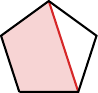}};
\node[inner sep=0pt] at (3.6,-1.7)
    {\includegraphics[scale=0.5]{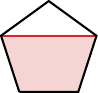}};
\draw [color=Blue,thick,-stealth] (4.15,0) -- (4.8,0);
\draw [color=Blue,thick,-stealth] (4.1,1.3) -- (4.95,0.25);
\draw [color=Blue,thick,-stealth] (4.1,-1.3) -- (4.95,-0.25);
\node[inner sep=0pt] at (5.45,0)
    {\includegraphics[scale=0.5]{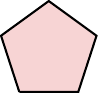}};
\end{tikzpicture}
\label{equ:5pt-Shared}
% \vspace{-10pt}
\eeq
To reduce the size of the figure, we have suppressed the solid/dashed distinction of the internal lines. We now see multiple shared functions between the different channels.

\vskip 2pt
A convenient way to visualize the relationship between the functions is to assign them to the geometry of an {\it associahedron}.
 To each vertex of the associahedron, we attach a complete triangulation of the kinematic polygon (and hence functions that are unique to the different channels), while the facets of the geometry correspond to partial triangulations (and hence shared functions).
For the case of the five-point function, we have 
\beq
\begin{tikzpicture}[baseline=(current  bounding  box.center),scale=.92]
\coordinate (1) at (-0.588*2,-0.809*2);
\coordinate (2) at (-0.951*2,0.309*2);
\coordinate (3) at (0,1*2);
\coordinate (4) at (0.951*2,0.309*2);
\coordinate (5) at (0.588*2,-0.809*2);
\coordinate (11) at (-0.588*2-.4,-0.809*2-.4);
\coordinate (22) at (-0.951*2-.6,0.309*2+.1);
\coordinate (33) at (0,1*2+.6);
\coordinate (44) at (0.951*2+.6,0.309*2+.1);
\coordinate (55) at (0.588*2+.4,-0.809*2-.4);
\coordinate (111) at (-0.588*2-.5-.3,-0.809*2+.95);
\coordinate (444) at (0.588*2+.5+.3,-0.809*2+.95);
\coordinate (222) at (-0.951*2-.7+1.3,0.309*2+.1+.95);
\coordinate (333) at (0.951*2+.7-1.3,0.309*2+.1+.95);
\coordinate (555) at (0,-0.809*2-.4);
\draw[black, line width=.6pt] (1) -- (2) -- (3) -- (4) -- (5) -- (1);
\node[inner sep=0pt] at (11)
    {\includegraphics[scale=0.5]{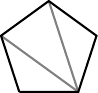}};
\node[inner sep=0pt] at (22)
    {\includegraphics[scale=0.5]{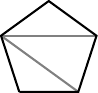}};
\node[inner sep=0pt] at (33)
    {\includegraphics[scale=0.5]{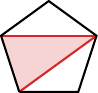}};
\node[inner sep=0pt] at (44)
    {\includegraphics[scale=0.5]{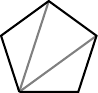}};
\node[inner sep=0pt] at (55)
    {\includegraphics[scale=0.5]{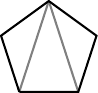}};
\node[inner sep=0pt] at (111)
    {\includegraphics[scale=0.45]{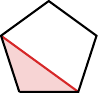}};
\node[inner sep=0pt] at (222)
    {\includegraphics[scale=0.45]{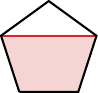}};
\node[inner sep=0pt] at (333)
    {\includegraphics[scale=0.45]{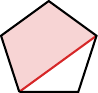}};
\node[inner sep=0pt] at (444)
    {\includegraphics[scale=0.45]{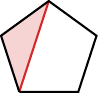}};
\node[inner sep=0pt] at (555)
    {\includegraphics[scale=0.45]{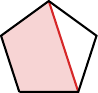}};
\node[inner sep=0pt] at (0,0)
    {\includegraphics[scale=0.5]{Figures/Shadings/Pentagon/5pt8}};
     \draw[-stealth, very thick,color=Red] (0.02,1*2) to[looseness=1.4,out=-85,in=-150] (0.75, 1.42);
          \draw[-stealth, very thick,color=Red] (-0.02,1*2) to[looseness=1.4,out=275,in=330] (-0.75, 1.42);
   \draw[-stealth, very thick,color=Green] (1.45,-0.8) to[looseness=1.1,out=170,in=80] (0.588*0.7,-0.809*2);
      \draw[-stealth, very thick,color=Orange] (-1.45,-0.8) to[looseness=1.1,out=10,in=100] (-0.588*0.7,-0.809*2);
           \draw[-stealth, very thick,color=Orange] (-1.63,-0.23) to[looseness=1.1,out=20,in=-60] (-1.3, 1.05);
             \draw[-stealth, very thick,color=Green] (1.63,-0.23) to[looseness=1.1,out=160,in=240] (1.3, 1.05);
   \draw[fill, color=regal] (3) circle (.75mm);
\draw [color=regal,ultra thick] (1) -- (2);
\draw [color=regal,ultra thick] (4) -- (5);
\draw [color=Blue,very thick,-stealth] (-0.951, 1.309) -- (-.4,.6);
\draw [color=Blue,very thick,-stealth] (0.951, 1.309) -- (.4,.6);
\draw [color=Blue,very thick,-stealth]  (0,-0.809*2) -- (0, -.8);
\end{tikzpicture}
\nonumber
\eeq
Shown here are the shadings appearing in \eqref{equ:5pt-Shared}. We have also indicated how the shadings evolve along the associahedron under the kinematic flow. Other shadings would appear if we were to take derivatives with respect to the other energies (and not just $k_1$). In general, we can think of the associahedron as being dressed with the basis functions and letters corresponding to the (partial) triangulations on its facets and vertices.
For the five-point function, there are 7 functions associated to each vertex, 4 shared functions at each edge and 1 shared function on the face of the pentagon~\cite{Companion}.

\vskip 4pt
\noindent
{\bf Outlook.} In this Letter, we introduced simple and universal rules underlying the apparent complexity of the differential equations for the FRW wavefunction of conformally coupled scalars.  These rules govern a flow in the boundary kinematic space and make no explicit reference to time evolution in the bulk spacetime.  
It is intriguing that the components of the construction appear naturally in the kinematic space of cosmological correlators, and are related to interesting combinatorial objects.

\vskip 2pt
We motivated this endeavor as the search for a viewpoint where time is a derived concept.
To view this as a true emergence of time, however, requires that this kinematic flow takes on a life of its own, by being derived from a different set of physical and mathematical principles. 
The mere existence of this structure is already a hint that a more elemental formulation exists, and we have seen glimpses of such hidden magic.
We hope that this work serves as inspiration for finding the deeper mathematical structure underlying cosmological correlations.

%\newpage
\vskip 8pt
\noindent
{\it Acknowledgements:} We are grateful for feedback and discussions to Ana Ach\'ucarro, Paolo Benincasa, Jan de Boer, Alessandra Caraceni, J.\,J.\,Carrasco, Xingang Chen, Claude Duhr, Alex Edison, Carolina Figueiredo, Dan Green, Thomas Grimm, Song He, Johannes Henn, Arno Hoefnagels,  Yu-tin Huang, Michael Jones, Manki Kim, Barak Kol, Chia-Kai Kuo, Daniel Longenecker, Manuel Loparco, Scott Melville, Sebastian Mizera, Matteo Parisi, Julio Parra-Martinez, Nic Pavao, Andrzej Pokraka, Oliver Schlotterer, Leonardo Senatore, Chia-Hsien Shen, John Stout, Bernd Sturmfels, Simon Telen, Jaroslav Trnka, Kamran Salehi Vaziri, Dong-Gang Wang and Alexander Zhiboedov. 

\vskip 2pt
NAH is supported by the US Department of Energy (DOE) under contract DE–SC0009988.
DB is supported by a Yushan Professorship at National Taiwan University funded by the Ministry of Education (Taiwan). AH is supported by DOE (HEP) Award DE-SC0011632 and by the Walter Burke Institute for Theoretical Physics. AJ is supported in part by DOE (HEP) Award DE-SC0009924. HL is supported by the Kavli Institute for Cosmological Physics at the University of Chicago through an endowment from the Kavli Foundation and its founder Fred Kavli. GLP is supported by a Rita-Levi Montalcini fellowship from the Italian Ministry of Universities and Research (MUR) and by INFN (IS GSS-Pi).

%\vspace{0.5cm}
%\clearpage
%\newpage
\appendix
\section{Equations for the Three-Site Chain}
\label{app:3site}

In this appendix, we will use the rules of the kinematic flow to derive the complete set of differential equations satisfied by the basis functions of the three-site chain.

\vskip 4pt
\noindent
{\bf Level 1:}
The equation of the wavefunction is given by the following evolutionary tree:
 \beq
 \begin{tikzpicture}[scale=0.7]
\node[inner sep=0pt] at (-0.4,0)
   {\includegraphics[scale=0.7]{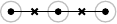}}; 
   \draw [color=gray,line width=0.5pt,-stealth] (0.8,0) -- (1.8,0);
\draw [color=gray,line width=0.5pt,-stealth] (0.8,0.15) -- (1.8,1.3);
\draw [color=gray,line width=0.5pt,-stealth] (0.8,-0.15) -- (1.8,-1.3);
\node[inner sep=0pt] at (3,1.4)
    {\includegraphics[scale=0.7]{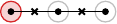}};
\node[inner sep=0pt] at (3,-1.4)
    {\includegraphics[scale=0.7]{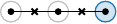}};  
    \node[inner sep=0pt] at (3,0)
    {\includegraphics[scale=0.7]{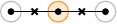}}; 
 \draw [color=gray,line width=0.5pt, dashed] (1.95,-1.65) -- (1.95,1.65) --  (4.05,1.65) -- (4.05,-1.65) -- (1.95,-1.65); 
  \node[below] at (3,-1.75)  {$\psi$};
       \draw [color=gray,line width=0.5pt,-stealth] (4.2,0.15) -- (5.05,0.5);
      \draw [color=gray,line width=0.5pt,-stealth] (4.2,0) -- (5.05,0);
         \draw [color=gray,line width=0.5pt,-stealth] (4.2,-0.15) -- (5.05,-0.5);
         \node[inner sep=0pt] at (6.15,0.58)
    {\includegraphics[scale=0.7]{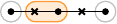}}; 
          \node[inner sep=0pt] at (6.15,0)
    {\includegraphics[scale=0.7]{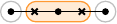}}; 
             \node[inner sep=0pt] at (6.15,-0.58)
    {\includegraphics[scale=0.7]{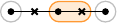}}; 
      \draw [color=gray,line width=0.5pt,-stealth] (4.2,1.4) -- (5.05,1.4);
    \draw [color=gray,line width=0.5pt,-stealth] (4.2,-1.4) -- (5.05,-1.4);
    \node[inner sep=0pt] at (6.15,1.4)
    {\includegraphics[scale=0.7]{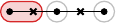}};
       \node[inner sep=0pt] at (6.15,-1.4)
    {\includegraphics[scale=0.7]{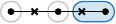}};
      \node[right] at (7.15,0.58)  {$Q_1$};
        \node[right] at (7.15,-0.58)  {$Q_3$};
           \node[right] at (7.15,0)  {$Q_2$};    
            \node[right] at (7.15,1.4)  {$F$};     
                 \node[right] at (7.15,-1.4)  {$\tilde F$}; 
\end{tikzpicture}
\nonumber
\eeq
We see that there are three ways in which the middle tube can grow to encircle the neighboring crosses. This produces the three source functions $Q_{1,2,3}$.  
From this tree, we infer the following differential
\beq
\boxed{
\begin{aligned}
\ud \psi
&=   \e\, \big[(\psi - F)\hspace{1pt}
\raisebox{-1.5pt}{\includegraphics[scale=0.7]{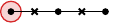}}
 +  
\big(\psi - {\textstyle \sum}Q_i \big) \hspace{-.5pt} \raisebox{-1.5pt}{\includegraphics[scale=0.7]{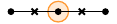}}
 \\
&\hspace{1.35cm} +    F\hspace{1pt}
\raisebox{-1.5pt}{\includegraphics[scale=0.7]{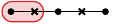}}
%%%%%%%%%%%%%%%%%%%%%%%%%%%%%%%%%%%%%%%%%%%%%
 \hspace{1.35cm} +    Q_{1}\hspace{-1pt}
\raisebox{-1.5pt}{\includegraphics[scale=0.7]{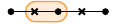}}
\\
&\hspace{0.83cm}(\psi - \tilde F)\hspace{1pt}
\raisebox{-1.5pt}{\includegraphics[scale=0.7]{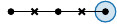}} \hspace{1.35cm} +    Q_{2}\hspace{-1pt}
\raisebox{-1.5pt}{\includegraphics[scale=0.7]{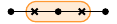}}
\\
&\hspace{1.35cm}  + \tilde F\hspace{1pt}
\raisebox{-1.5pt}{\includegraphics[scale=0.7]{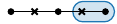}}
 \hspace{1.35cm}  +     Q_3\hspace{-1pt}
\raisebox{-1.5pt}{\includegraphics[scale=0.7]{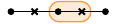}}  \big] 
\end{aligned}}
\nonumber
\eeq
which is similar to (\ref{eq:twositedpsi}) for the two-site chain.

\vskip 4pt
\noindent
{\bf Level 2:} The differential of $\psi$ involved five source functions, associated to the following graph tubings:
\beq
\begin{aligned}
\quad & F\,\   \raisebox{-1pt}{\includegraphics[scale=0.7]{Figures/Tubings/three/tree/threeF}}   
\quad && Q_1\,\ \raisebox{-1pt}{\includegraphics[scale=0.7]{Figures/Tubings/three/tree/threeQ1}}  
\\[-1pt]
& \tilde F\,\  \raisebox{-1pt}{\includegraphics[scale=0.7]{Figures/Tubings/three/tree/threeFt}}  
&& Q_2\,\ \raisebox{-1pt}{\includegraphics[scale=0.7]{Figures/Tubings/three/tree/threeQ2}} 
\\
& 
&& Q_3\,\  \raisebox{-1pt}{\includegraphics[scale=0.7]{Figures/Tubings/three/tree/threeQ3}} 
\end{aligned}
\nonumber
\eeq
We will now predict the differentials of these functions.

\vskip 4pt
\noindent
$\bullet$\
The differential of the function $F$ is predicted by 
 \beq
 \begin{tikzpicture}[scale=0.7]
\node[inner sep=0pt] at (-0.4,0)
   {\includegraphics[scale=0.7]{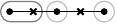}}; 
   \draw [color=gray,line width=0.5,-stealth] (0.8,0) -- (1.8,0);
\draw [color=gray,line width=0.5,-stealth] (0.8,0.15) -- (1.8,1.3);
\draw [color=gray,line width=0.5,-stealth] (0.8,-0.15) -- (1.8,-1.3);
\node[inner sep=0pt] at (3,1.4)
    {\includegraphics[scale=0.7]{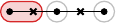}};
\node[inner sep=0pt] at (3,-1.4)
    {\includegraphics[scale=0.7]{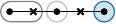}};  
    \node[inner sep=0pt] at (3,0)
    {\includegraphics[scale=0.7]{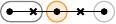}}; 
 \draw [color=gray,line width=0.5,line width=0.5pt, dashed] (1.95,-1.65) -- (1.95,1.65) --  (4.05,1.65) -- (4.05,-1.65) -- (1.95,-1.65); 
  \node[below] at (3,-1.75)  {$F$};
       \draw [color=gray,line width=0.5,-stealth] (4.2,0.15) -- (5.05,0.5);
      \draw [color=gray,line width=0.5,-stealth] (4.2,0) -- (5.05,0);
         \draw [color=gray,line width=0.5,-stealth] (4.2,-0.15) -- (5.05,-0.5);
         \node[inner sep=0pt] at (6.15,0.55)
    {\includegraphics[scale=0.7]{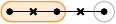}}; 
          \node[inner sep=0pt] at (6.15,0)
    {\includegraphics[scale=0.7]{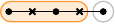}}; 
             \node[inner sep=0pt] at (6.15,-0.55)
    {\includegraphics[scale=0.7]{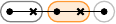}}; 
    \draw [color=gray,line width=0.5,-stealth] (4.2,-1.4) -- (5.05,-1.4);
       \node[inner sep=0pt] at (6.15,-1.4)
    {\includegraphics[scale=0.7]{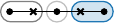}};
    
      \node[right] at (7.15,0.55)  {$q_1$};
        \node[right] at (7.15,-0.55)  {$q_3$};
           \node[right] at (7.15,0)  {$q_2$};
                 \node[right] at (7.15,-1.4)  {$f$}; 
\end{tikzpicture}
\nonumber
\eeq
Note that
the functions $q_1$ and $q_2$ are created by mergers, while the functions $q_3$ and $f$ correspond to disconnected tubings.
The differential of $F$ then is
\beq
\boxed{
\begin{aligned}
\ud F
&=    \e\, \big[ F\hspace{1pt}
\raisebox{-1.5pt}{\includegraphics[scale=0.7]{Figures/Tubings/three/letters/LLam.pdf}}
   \,+ \,
\big(F - {\textstyle \sum} q_{i}\big)
\raisebox{-1.5pt}{\includegraphics[scale=0.7]{Figures/Tubings/three/letters/LLbpp.pdf}}
  \\
 &
 \hspace{4cm} +  q_1\hskip 1pt \raisebox{-1.5pt}{\includegraphics[scale=0.7]{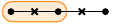}}
 \\[2pt]
&\hspace{0.9cm} (F-f) 
\raisebox{-1.5pt}{\includegraphics[scale=0.7]{Figures/Tubings/three/letters/LLcp.pdf}}
\hspace{0.55cm}
  +   q_2\hskip 1pt \raisebox{-1.5pt}{\includegraphics[scale=0.7]{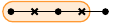}}
 \\
&\hspace{1.4cm} +f  \hskip 1pt \raisebox{-1.5pt}{\includegraphics[scale=0.7]{Figures/Tubings/three/letters/LLcm.pdf}}
\hspace{0.55cm}
 +     q_3\hskip 1pt \raisebox{-1.5pt}{\includegraphics[scale=0.7]{Figures/Tubings/three/letters/LLbpm.pdf}}
\ \big]
\end{aligned}}
\nonumber
\eeq

\vskip 4pt
\noindent
$\bullet$\
The tree for the function $Q_1$ was presented in the main text. We reproduce it here for completeness:
 \beq
 \begin{tikzpicture}[scale=0.7]
\node[inner sep=0pt] at (0.6,0)
    {\includegraphics[scale=0.7]{Figures/Tubings/three/tree/threeQ1}}; 
\draw [color=gray,-stealth] (1.8,0) -- (2.6,0);
\draw [color=gray,-stealth] (1.8,0.15) -- (2.6,0.6);
\draw [color=gray,-stealth] (1.8,-0.15) -- (2.6,-0.6);
\node[inner sep=0pt] at (3.8,0.75)
    {\includegraphics[scale=0.7]{Figures/Tubings/three/tree/threeQ1a}};
\node[inner sep=0pt] at (3.8,0)
    {\includegraphics[scale=0.7]{Figures/Tubings/three/tree/threeQ1b}}; 
\node[inner sep=0pt] at (3.8,-0.75)
    {\includegraphics[scale=0.7]{Figures/Tubings/three/tree/threeQ1c}}; 
\draw [color=gray,line width=0.5pt, dashed] (2.75,-1.03) -- (2.75,1.02) --  (4.85,1.02) -- (4.85,-1.03) -- (2.75,-1.03);  
\node[below] at (3.8,-1.03)  {$Q_1$};   
\draw [color=gray,-stealth] (5.,0.75) -- (5.8,0.75);
\draw [color=gray,-stealth] (5,-0.75) -- (5.8,-0.75);
\node[inner sep=0pt] at (7.,0.75)
    {\includegraphics[scale=0.7]{Figures/Tubings/three/tree/threeq1ab}};
            \node[above] at (7,1.02)  {$q_1$}; 
\node[inner sep=0pt] at (7.,-0.75)
    {\includegraphics[scale=0.7]{Figures/Tubings/three/tree/threeqt3c}};
        \node[below] at (7,-1.03)  {$\tilde q_3$}; 
\draw [color=gray,-stealth] (8.2,-0.75) -- (9.,-0.75);
\node[inner sep=0pt] at (10.2,-0.75)
    {\includegraphics[scale=0.7]{Figures/Tubings/three/tree/threeqt2bc}};
    \node[below] at (10.2,-1.03)  {$-\tilde q_2$}; 
\end{tikzpicture}
\nonumber
\eeq
The top branch shows growth and merger, while the bottom branch contains the first instance of the absorption phenomenon. The differential of the function $Q_1$ then is 
\beq
\boxed{
\begin{aligned}
\ud Q_1
&=     \e\, \big[
(Q_1 -q_1)
    \,\hskip .5pt\raisebox{-1pt}{\includegraphics[scale=0.7]{Figures/Tubings/three/letters/LLap}} 
+ q_1\,\raisebox{-1pt}{\includegraphics[scale=0.7]{Figures/Tubings/three/letters/LLabpr}} 
 \\ 
&\hspace{1.5cm} +    
 Q_{1} \raisebox{-1pt}{\includegraphics[scale=0.7]{Figures/Tubings/three/letters/LLbmp}} \\
&\hspace{0.48cm}+(Q_1 - \tilde q_{3})   \raisebox{-1pt}{\includegraphics[scale=0.7]{Figures/Tubings/three/letters/LLcp}}  \\
&\hspace{0.48cm}+ \,\, (\tilde q_3  + \tilde q_2)
  \raisebox{-1pt}{\includegraphics[scale=0.7]{Figures/Tubings/three/letters/LLcm}}    -   \tilde q_2
  \,\raisebox{-1pt}{\includegraphics[scale=0.7]{Figures/Tubings/three/letters/LLbcmb}} \, \big]
\end{aligned}}
\nonumber
\eeq
\vskip 4pt
\noindent
$\bullet$\ By symmetry, the differential for the function $Q_3$ is equivalent to that of the function $Q_1$, and therefore won't be shown explicitly.

\vskip 4pt
\noindent
$\bullet$\ 
The tree for the function $Q_2$ is 
 \beq
 \begin{tikzpicture}[scale=0.7]
\node[inner sep=0pt] at (0.6,0)
    {\includegraphics[scale=0.7]{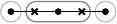}}; 
\draw [color=gray,-stealth] (1.8,0) -- (2.6,0);
\draw [color=gray,-stealth] (1.8,0.15) -- (2.6,0.6);
\draw [color=gray,-stealth] (1.8,-0.15) -- (2.6,-0.6);
\node[inner sep=0pt] at (3.8,0.75)
    {\includegraphics[scale=0.7]{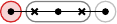}};
\node[inner sep=0pt] at (3.8,0)
    {\includegraphics[scale=0.7]{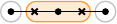}}; 
\node[inner sep=0pt] at (3.8,-0.75)
    {\includegraphics[scale=0.7]{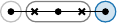}}; 
\draw [color=gray,line width=0.5pt, dashed] (2.75,-1.03) -- (2.75,1.02) --  (4.85,1.02) -- (4.85,-1.03) -- (2.75,-1.03);  
\node[below] at (3.8,-1.03)  {$Q_2$};   
\draw [color=gray,-stealth] (5.,0.75) -- (5.8,0.75);
\draw [color=gray,-stealth] (5,-0.75) -- (5.8,-0.75);
\node[inner sep=0pt] at (7.,0.75)
    {\includegraphics[scale=0.7]{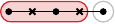}};
\node[inner sep=0pt] at (7.,-0.75)
    {\includegraphics[scale=0.7]{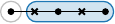}};
                    \node[right] at (8,0.75)  {$q_2$};                       
                 \node[right] at (8,-0.75)  {$\tilde q_2$}; 
\end{tikzpicture}
\nonumber
\eeq
where the functions $q_2$ and $\tilde q_2$ are created by the standard growth and merger of the tubes. The differential of $Q_2$ then is
\beq
\boxed{\begin{aligned}
\ud Q_2
&=     \e\, \big[
(Q_2 -q_2)
    \,\hskip .5pt\raisebox{-1pt}{\includegraphics[scale=0.7]{Figures/Tubings/three/letters/LLap}} 
+ q_2\,\raisebox{-1pt}{\includegraphics[scale=0.7]{Figures/Tubings/three/letters/LLabmr}} 
 \\ 
&\hspace{1.5cm} +    
 Q_{2} \raisebox{-1pt}{\includegraphics[scale=0.7]{Figures/Tubings/three/letters/LLbmm}} \\
&\hspace{0.48cm}+(Q_2 - \tilde q_{2})   \raisebox{-1pt}{\includegraphics[scale=0.7]{Figures/Tubings/three/letters/LLcp}} +  \tilde q_2
 \, \raisebox{-1pt}{\includegraphics[scale=0.7]{Figures/Tubings/three/letters/LLbcmb}} \, \big]
\end{aligned}}
\nonumber
\eeq

\vskip 4pt
\noindent
 {\bf Level 3:} The differentials at Level~2 produced 7 source functions corresponding to the following graph tubings:
\begin{align}
\begin{split}
f\,  \raisebox{-3pt}{\includegraphics[scale=0.7]{Figures/Tubings/three/tree/threeff}}    \hspace{1.05cm}
&q_1\,  \raisebox{-3pt}{\includegraphics[scale=0.7]{Figures/Tubings/three/tree/threeqq1}}  \hspace{1cm} 
\tilde q_1\, \raisebox{-3pt}{\includegraphics[scale=0.7]{Figures/Tubings/three/tree/threeqqt1}} \\
&q_2\, \raisebox{-3pt}{\includegraphics[scale=0.7]{Figures/Tubings/three/tree/threeqq2}} \hspace{1cm} 
\tilde q_2\, \raisebox{-3pt}{\includegraphics[scale=0.7]{Figures/Tubings/three/tree/threeqqt2}} \\
&q_3\,\raisebox{-3pt}{\includegraphics[scale=0.7]{Figures/Tubings/three/tree/threeqq3}}  \hspace{1.cm}
 \tilde q_3\, \raisebox{-3pt}{\includegraphics[scale=0.7]{Figures/Tubings/three/tree/threeqqt3}}
\end{split}
\nonumber
\end{align}

\noindent
$\bullet$\ The tree for the function $f$ is
 \beq
 \begin{tikzpicture}[scale=0.7]
  %\node at (-1.9,0)  {$f$\,:};
\node[inner sep=0pt] at (-0.4,0)
   {\includegraphics[scale=0.7]{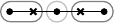}}; 
   \draw [color=gray,line width=0.5,-stealth] (0.8,0) -- (1.8,0);
\draw [color=gray,line width=0.5,-stealth] (0.8,0.15) -- (1.8,1.3);
\draw [color=gray,line width=0.5,-stealth] (0.8,-0.15) -- (1.8,-1.3);
\node[inner sep=0pt] at (3,1.4)
    {\includegraphics[scale=0.7]{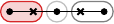}};
\node[inner sep=0pt] at (3,-1.4)
    {\includegraphics[scale=0.7]{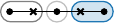}};  
    \node[inner sep=0pt] at (3,0)
    {\includegraphics[scale=0.7]{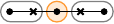}}; 
    %\node at (6.4,-0.7)  {$\tilde F$};
 \draw [color=gray,line width=0.5,line width=0.5pt, dashed] (1.95,-1.65) -- (1.95,1.65) --  (4.05,1.65) -- (4.05,-1.65) -- (1.95,-1.65); 
  \node[below] at (3,-1.75)  {$f$};
  
       \draw [color=gray,line width=0.5,-stealth] (4.2,0.15) -- (5.05,0.5);
      \draw [color=gray,line width=0.5,-stealth] (4.2,0) -- (5.05,0);
         \draw [color=gray,line width=0.5,-stealth] (4.2,-0.15) -- (5.05,-0.5);
         \node[inner sep=0pt] at (6.15,0.55)
    {\includegraphics[scale=0.7]{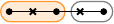}}; 
          \node[inner sep=0pt] at (6.15,-0.55)
    {\includegraphics[scale=0.7]{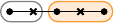}}; 
             \node[inner sep=0pt] at (6.15,0)
    {\includegraphics[scale=0.7]{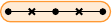}}; 
    
          \node[right] at (7.15,0.55)  {$g$};
        \node[right] at (7.15,0)  {$Z$};
           \node[right] at (7.15,-0.55)  {$\tilde g$};      
\end{tikzpicture}
\nonumber
\eeq
Note that the function $Z$ is created by the merger of three tubes.
The differential of $f$ then is
\beq
\boxed{
\begin{aligned}
\ud f
=    \e\, \big[ f\hspace{1pt}
\raisebox{-1.5pt}{\includegraphics[scale=0.7]{Figures/Tubings/three/letters/LLam.pdf}}
   \,+ \,
f  \hskip 1pt &\raisebox{-1.5pt}{\includegraphics[scale=0.7]{Figures/Tubings/three/letters/LLcm.pdf}}
  \\
  \big(F - g - \tilde g - Z\big)
&\raisebox{-1.5pt}{\includegraphics[scale=0.7]{Figures/Tubings/three/letters/LLbpp.pdf}}\\
+\  g\hskip 1pt &\raisebox{-1.5pt}{\includegraphics[scale=0.7]{Figures/Tubings/three/letters/LLabpo.pdf}}
 \\[2pt]
  + \  \tilde g\hskip 1pt &\raisebox{-1.5pt}{\includegraphics[scale=0.7]{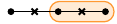}}
 \\
+\ Z\hskip 2pt &\,\raisebox{-1.5pt}{\includegraphics[scale=0.7]{Figures/Tubings/three/ff/threeZb}}
\ \big]
\end{aligned}}
\nonumber
\eeq

\vskip 4pt
\noindent
$\bullet$\ The tree for the function $q_1$ is
 \beq
 \begin{tikzpicture}[scale=0.7]
\node[inner sep=0pt] at (0.6,0)
    {\includegraphics[scale=0.7]{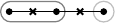}}; 
\draw [color=gray,-stealth] (1.8,0.15) -- (2.6,0.6);
\draw [color=gray,-stealth] (1.8,-0.15) -- (2.6,-0.6);
\node[inner sep=0pt] at (3.8,0.75)
    {\includegraphics[scale=0.7]{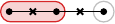}};
\node[inner sep=0pt] at (3.8,-0.75)
    {\includegraphics[scale=0.7]{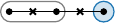}}; 
\draw [color=gray,line width=0.5pt, dashed] (2.75,-1.03) -- (2.75,1.02) --  (4.85,1.02) -- (4.85,-1.03) -- (2.75,-1.03);  
\node[above] at (3.8,1.03)  {$2q_1$};   
\node[below] at (3.8,-1.03)  {$q_1$};   
\draw [color=gray,-stealth] (5,-0.75) -- (5.8,-0.75);
\node[inner sep=0pt] at (7.,-0.75)
    {\includegraphics[scale=0.7]{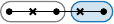}};
        \node[below] at (7,-1.03)  {$g$}; 
\draw [color=gray,-stealth] (8.2,-0.75) -- (9.,-0.75);
\node[inner sep=0pt] at (10.2,-0.75)
    {\includegraphics[scale=0.7]{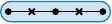}};
    \node[below] at (10.2,-1.03)  {$-Z$}; 
\end{tikzpicture}
\nonumber
\eeq
where we have an instance of growth and absorption in the bottom branch. The assigned function in the top branch comes with a factor of 2 since the activated tube encloses 2 vertices.
The differential of $q_1$ then is
\beq
\boxed{
\begin{aligned}
\ud q_1
=    \e\, \big[ 2 q_1\hspace{1pt}
\raisebox{-1.5pt}{\includegraphics[scale=0.7]{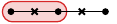}}
   \,+ \,
(q_1-g)  \hskip 1pt &\raisebox{-1.5pt}{\includegraphics[scale=0.7]{Figures/Tubings/three/letters/LLcp.pdf}}
  \\
  \big(g + Z\big)
&\raisebox{-1.5pt}{\includegraphics[scale=0.7]{Figures/Tubings/three/letters/LLcm.pdf}}\\
- Z\hskip 2pt &\raisebox{-1.5pt}{\includegraphics[scale=0.7]{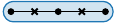}}
\ \big]
\end{aligned}}
\nonumber
\eeq

\vskip 4pt
\noindent
$\bullet$\ The tree for the function $q_2$ is
 \beq
 \begin{tikzpicture}[scale=0.7]
\node[inner sep=0pt] at (0.6,0)
    {\includegraphics[scale=0.7]{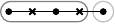}}; 
\draw [color=gray,-stealth] (1.8,0.15) -- (2.6,0.6);
\draw [color=gray,-stealth] (1.8,-0.15) -- (2.6,-0.6);
\node[inner sep=0pt] at (3.8,0.75)
    {\includegraphics[scale=0.7]{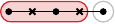}};
\node[inner sep=0pt] at (3.8,-0.75)
    {\includegraphics[scale=0.7]{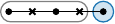}}; 
\draw [color=gray,line width=0.5pt, dashed] (2.75,-1.03) -- (2.75,1.02) --  (4.85,1.02) -- (4.85,-1.03) -- (2.75,-1.03);  
\node[above] at (3.8,1.03)  {$2q_2$};   
\node[below] at (3.8,-1.03)  {$q_2$};   
\draw [color=gray,-stealth] (5,-0.75) -- (5.8,-0.75);
\node[inner sep=0pt] at (7.,-0.75)
    {\includegraphics[scale=0.7]{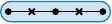}};
        \node[below] at (7,-1.03)  {$Z$}; 
\end{tikzpicture}
\nonumber
\eeq
so that the differential is
\beq
\boxed{
\begin{aligned}
\ud q_2
=    \e\, \big[\, 2 q_2\hspace{1pt}
\raisebox{-1.5pt}{\includegraphics[scale=0.7]{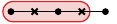}}
   \,+ \,
(q_2-Z)  \hskip 1pt &\raisebox{-1.5pt}{\includegraphics[scale=0.7]{Figures/Tubings/three/letters/LLcp.pdf}}\\
+\ Z\hskip 2pt &\raisebox{-1.5pt}{\includegraphics[scale=0.7]{Figures/Tubings/three/letters/LLabcb.pdf}}
\ \big]
\end{aligned}}
\nonumber
\eeq

\vskip 4pt
\noindent
$\bullet$\ The tree for the function $q_3$ is
 \beq
 \begin{tikzpicture}[scale=0.7]
\node[inner sep=0pt] at (0.6,0)
    {\includegraphics[scale=0.7]{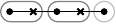}}; 
\draw [color=gray,-stealth] (1.8,0) -- (2.6,0);
\draw [color=gray,-stealth] (1.8,0.15) -- (2.6,0.6);
\draw [color=gray,-stealth] (1.8,-0.15) -- (2.6,-0.6);
\node[inner sep=0pt] at (3.8,0.75)
    {\includegraphics[scale=0.7]{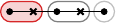}};
\node[inner sep=0pt] at (3.8,0)
    {\includegraphics[scale=0.7]{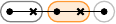}}; 
\node[inner sep=0pt] at (3.8,-0.75)
    {\includegraphics[scale=0.7]{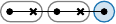}}; 
\draw [color=gray,line width=0.5pt, dashed] (2.75,-1.03) -- (2.75,1.02) --  (4.85,1.02) -- (4.85,-1.03) -- (2.75,-1.03);  
\node[below] at (3.8,-1.03)  {$q_3$};   
\draw [color=gray,-stealth] (5.,0.75) -- (5.8,0.75);
\draw [color=gray,-stealth] (5,-0.75) -- (5.8,-0.75);
\node[inner sep=0pt] at (7.,0.75)
    {\includegraphics[scale=0.7]{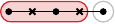}};
\node[inner sep=0pt] at (7.,-0.75)
    {\includegraphics[scale=0.7]{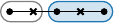}};
                    \node[right] at (8,0.75)  {$(-q_2)$};                       
                 \node[right] at (8,-0.75)  {$\tilde g$}; 
\end{tikzpicture}
\nonumber
\eeq
In the top branch, the left tube is first activated and then absorbs its neighboring tube. 
The remaining features of the tree are the standard activation, growth and merger.
The differential of $q_3$ then is
\beq
\boxed{\begin{aligned}
\ud q_3
=     \e\, \big[
(q_3 +q_2)
    \,&\raisebox{-1pt}{\includegraphics[scale=0.7]{Figures/Tubings/three/letters/LLam}} 
- q_2\,\raisebox{-1pt}{\includegraphics[scale=0.7]{Figures/Tubings/three/letters/LLabmr}} 
 \\ 
 +\,    
 q_3 &\raisebox{-1pt}{\includegraphics[scale=0.7]{Figures/Tubings/three/letters/LLbpm}} \\
+\,(q_3 - \tilde g)   &\raisebox{-1pt}{\includegraphics[scale=0.7]{Figures/Tubings/three/letters/LLcp}} +  \tilde g\,
  \raisebox{-1pt}{\includegraphics[scale=0.7]{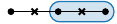}} \, \big]
\end{aligned}}
\nonumber
\eeq

\vskip 4pt
\noindent
$\bullet$\ 
The differentials for $\tilde q_{1,2,3}$ are related to those of $q_{1,2,3}$ by symmetry and are therefore not shown explicitly.

\vskip 4pt
\noindent
 {\bf Level 4:} At Level~$3$, we obtained three new source functions corresponding to the following graph tubings: 
\begin{align}
\begin{split}
&g\, \raisebox{-1pt}{\includegraphics[scale=0.7]{Figures/Tubings/three/tree/threeg}}
\hspace{1cm}
\tilde g\, \raisebox{-1pt}{\includegraphics[scale=0.7]{Figures/Tubings/three/tree/threegt}}
 \hspace{1cm}
Z\, \raisebox{-1pt}{\includegraphics[scale=0.7]{Figures/Tubings/three/tree/threeZ}}
\end{split}
\nonumber
\end{align}
To complete our analysis, we consider their differentials.

\vskip 4pt
\noindent
$\bullet$\ The tree for the function $g$ is
 \beq
 \begin{tikzpicture}[scale=0.7]
\node[inner sep=0pt] at (0.6,0)
    {\includegraphics[scale=0.7]{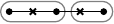}}; 
\draw [color=gray,-stealth] (1.8,0.15) -- (2.6,0.6);
\draw [color=gray,-stealth] (1.8,-0.15) -- (2.6,-0.6);
\node[inner sep=0pt] at (3.8,0.75)
    {\includegraphics[scale=0.7]{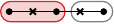}};
\node[inner sep=0pt] at (3.8,-0.75)
    {\includegraphics[scale=0.7]{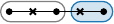}}; 
\draw [color=gray,line width=0.5pt, dashed] (2.75,-1.03) -- (2.75,1.02) --  (4.85,1.02) -- (4.85,-1.03) -- (2.75,-1.03);  
\node[above] at (3.8,1.03)  {$2g$};   
\node[below] at (3.8,-1.03)  {$g$};   
\draw [color=gray,-stealth] (5,-0.75) -- (5.8,-0.75);
\node[inner sep=0pt] at (7.,-0.75)
    {\includegraphics[scale=0.7]{Figures/Tubings/three/qq2/threeZc}};
        \node[below] at (7,-1.03)  {$-Z$}; 
\end{tikzpicture}
\nonumber
\eeq
which is similar to the tree for $q_2$, expect that the function $Z$ is created by absorption (rather than merger) and therefore comes with a minus sign.
The differential of $g$ then is
\beq
\boxed{
\begin{aligned}
\ud g
=    \e\, \big[ \, 2 g\hspace{1pt}
\raisebox{-1.5pt}{\includegraphics[scale=0.7]{Figures/Tubings/three/letters/LLabmr.pdf}}
   \,+ \,
(g+Z)  \hskip 1pt &\raisebox{-1.5pt}{\includegraphics[scale=0.7]{Figures/Tubings/three/letters/LLcp.pdf}}\\
-\ Z\hskip 2pt &\raisebox{-1.5pt}{\includegraphics[scale=0.7]{Figures/Tubings/three/letters/LLabcb.pdf}}
\ \big]
\end{aligned}}
\nonumber
\eeq

\vskip 4pt
\noindent
$\bullet$\ The differential for the function $\tilde g$ is related to that of the function $g$ by symmetry and therefore is not shown explicitly.

\vskip 4pt
\noindent
$\bullet$\ 
Finally, the tree for the function $Z$ is
 \beq
 \begin{tikzpicture}[scale=0.7]
\node[inner sep=0pt] at (-0.4,0)
   {\includegraphics[scale=0.7]{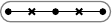}}; 
   \draw [color=gray,line width=0.5,-stealth] (0.8,0) -- (1.8,0);
    \node[inner sep=0pt] at (3,0)
    {\includegraphics[scale=0.7]{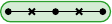}}; 
 \node at (4.4,0)  {$3Z$}; 
\end{tikzpicture}
\nonumber
\eeq
The tube is simply activated and the assigned coefficient function is 3 times the original function (since the activated tube contains 3 vertices). The differential of $Z$ then is
\beq
\boxed{
\ud Z = 3\e\, Z\,    \raisebox{-1.5pt}{\includegraphics[scale=0.7]{Figures/Tubings/three/Z/threeZgr}}}
\nonumber
\eeq
We have seen that all equations for the three-site chain are predicted by the simple rules of the kinematic flow. In fact, as we show in our companion paper~\cite{Companion}, the same rules underly the equations for arbitrary tree graphs.
\newpage
\bibliography{Flow-Letter-Refs}

%apsrev4-2.bst 2019-01-14 (MD) hand-edited version of apsrev4-1.bst
%Control: key (0)
%Control: author (8) initials jnrlst
%Control: editor formatted (1) identically to author
%Control: production of article title (0) allowed
%Control: page (0) single
%Control: year (1) truncated
%Control: production of eprint (0) enabled
\begin{thebibliography}{31}%
\makeatletter
\providecommand \@ifxundefined [1]{%
 \@ifx{#1\undefined}
}%
\providecommand \@ifnum [1]{%
 \ifnum #1\expandafter \@firstoftwo
 \else \expandafter \@secondoftwo
 \fi
}%
\providecommand \@ifx [1]{%
 \ifx #1\expandafter \@firstoftwo
 \else \expandafter \@secondoftwo
 \fi
}%
\providecommand \natexlab [1]{#1}%
\providecommand \enquote  [1]{``#1''}%
\providecommand \bibnamefont  [1]{#1}%
\providecommand \bibfnamefont [1]{#1}%
\providecommand \citenamefont [1]{#1}%
\providecommand \href@noop [0]{\@secondoftwo}%
\providecommand \href [0]{\begingroup \@sanitize@url \@href}%
\providecommand \@href[1]{\@@startlink{#1}\@@href}%
\providecommand \@@href[1]{\endgroup#1\@@endlink}%
\providecommand \@sanitize@url [0]{\catcode `\\12\catcode `\$12\catcode
  `\&12\catcode `\#12\catcode `\^12\catcode `\_12\catcode `\%12\relax}%
\providecommand \@@startlink[1]{}%
\providecommand \@@endlink[0]{}%
\providecommand \url  [0]{\begingroup\@sanitize@url \@url }%
\providecommand \@url [1]{\endgroup\@href {#1}{\urlprefix }}%
\providecommand \urlprefix  [0]{URL }%
\providecommand \Eprint [0]{\href }%
\providecommand \doibase [0]{https://doi.org/}%
\providecommand \selectlanguage [0]{\@gobble}%
\providecommand \bibinfo  [0]{\@secondoftwo}%
\providecommand \bibfield  [0]{\@secondoftwo}%
\providecommand \translation [1]{[#1]}%
\providecommand \BibitemOpen [0]{}%
\providecommand \bibitemStop [0]{}%
\providecommand \bibitemNoStop [0]{.\EOS\space}%
\providecommand \EOS [0]{\spacefactor3000\relax}%
\providecommand \BibitemShut  [1]{\csname bibitem#1\endcsname}%
\let\auto@bib@innerbib\@empty
%</preamble>
\bibitem [{\citenamefont {Baumann}\ \emph {et~al.}()\citenamefont {Baumann},
  \citenamefont {Green}, \citenamefont {Joyce}, \citenamefont {Pajer},
  \citenamefont {Pimentel}, \citenamefont {Sleight},\ and\ \citenamefont
  {Taronna}}]{Baumann:2022jpr}%
  \BibitemOpen
  \bibfield  {author} {\bibinfo {author} {\bibfnamefont {D.}~\bibnamefont
  {Baumann}}, \bibinfo {author} {\bibfnamefont {D.}~\bibnamefont {Green}},
  \bibinfo {author} {\bibfnamefont {A.}~\bibnamefont {Joyce}}, \bibinfo
  {author} {\bibfnamefont {E.}~\bibnamefont {Pajer}}, \bibinfo {author}
  {\bibfnamefont {G.}~\bibnamefont {Pimentel}}, \bibinfo {author}
  {\bibfnamefont {C.}~\bibnamefont {Sleight}},\ and\ \bibinfo {author}
  {\bibfnamefont {M.}~\bibnamefont {Taronna}},\ }\bibfield  {title} {\bibinfo
  {title} {{Snowmass White Paper: The Cosmological Bootstrap}},\ }in\
  \href@noop {} {\emph {\bibinfo {booktitle} {{Snowmass Summer Study}}}},\
  \Eprint {https://arxiv.org/abs/2203.08121} {arXiv:2203.08121 [hep-th]}
  \BibitemShut {NoStop}%
\bibitem [{\citenamefont {Maldacena}\ and\ \citenamefont
  {Pimentel}(2011)}]{Maldacena:2011nz}%
  \BibitemOpen
  \bibfield  {author} {\bibinfo {author} {\bibfnamefont {J.}~\bibnamefont
  {Maldacena}}\ and\ \bibinfo {author} {\bibfnamefont {G.}~\bibnamefont
  {Pimentel}},\ }\bibfield  {title} {\bibinfo {title} {{On Graviton
  Non-Gaussianities during Inflation}},\ }\href
  {https://doi.org/10.1007/JHEP09(2011)045} {\bibfield  {journal} {\bibinfo
  {journal} {JHEP}\ }\textbf {\bibinfo {volume} {09}},\ \bibinfo {pages}
  {045}},\ \Eprint {https://arxiv.org/abs/1104.2846} {arXiv:1104.2846 [hep-th]}
  \BibitemShut {NoStop}%
\bibitem [{\citenamefont {Mata}\ \emph {et~al.}(2013)\citenamefont {Mata},
  \citenamefont {Raju},\ and\ \citenamefont {Trivedi}}]{Mata:2012bx}%
  \BibitemOpen
  \bibfield  {author} {\bibinfo {author} {\bibfnamefont {I.}~\bibnamefont
  {Mata}}, \bibinfo {author} {\bibfnamefont {S.}~\bibnamefont {Raju}},\ and\
  \bibinfo {author} {\bibfnamefont {S.}~\bibnamefont {Trivedi}},\ }\bibfield
  {title} {\bibinfo {title} {{CMB from CFT}},\ }\href
  {https://doi.org/10.1007/JHEP07(2013)015} {\bibfield  {journal} {\bibinfo
  {journal} {JHEP}\ }\textbf {\bibinfo {volume} {07}},\ \bibinfo {pages}
  {015}},\ \Eprint {https://arxiv.org/abs/1211.5482} {arXiv:1211.5482 [hep-th]}
  \BibitemShut {NoStop}%
\bibitem [{\citenamefont {McFadden}\ and\ \citenamefont
  {Skenderis}(2010)}]{McFadden:2009fg}%
  \BibitemOpen
  \bibfield  {author} {\bibinfo {author} {\bibfnamefont {P.}~\bibnamefont
  {McFadden}}\ and\ \bibinfo {author} {\bibfnamefont {K.}~\bibnamefont
  {Skenderis}},\ }\bibfield  {title} {\bibinfo {title} {{Holography for
  Cosmology}},\ }\href {https://doi.org/10.1103/PhysRevD.81.021301} {\bibfield
  {journal} {\bibinfo  {journal} {Phys. Rev. D}\ }\textbf {\bibinfo {volume}
  {81}},\ \bibinfo {pages} {021301} (\bibinfo {year} {2010})},\ \Eprint
  {https://arxiv.org/abs/0907.5542} {arXiv:0907.5542 [hep-th]} \BibitemShut
  {NoStop}%
\bibitem [{\citenamefont {Bzowski}\ \emph {et~al.}(2014)\citenamefont
  {Bzowski}, \citenamefont {McFadden},\ and\ \citenamefont
  {Skenderis}}]{Bzowski:2013sza}%
  \BibitemOpen
  \bibfield  {author} {\bibinfo {author} {\bibfnamefont {A.}~\bibnamefont
  {Bzowski}}, \bibinfo {author} {\bibfnamefont {P.}~\bibnamefont {McFadden}},\
  and\ \bibinfo {author} {\bibfnamefont {K.}~\bibnamefont {Skenderis}},\
  }\bibfield  {title} {\bibinfo {title} {{Implications of Conformal Invariance
  in Momentum Space}},\ }\href {https://doi.org/10.1007/JHEP03(2014)111}
  {\bibfield  {journal} {\bibinfo  {journal} {JHEP}\ }\textbf {\bibinfo
  {volume} {03}},\ \bibinfo {pages} {111}},\ \Eprint
  {https://arxiv.org/abs/1304.7760} {arXiv:1304.7760 [hep-th]} \BibitemShut
  {NoStop}%
\bibitem [{\citenamefont {Arkani-Hamed}\ and\ \citenamefont
  {Maldacena}(2015)}]{Arkani-Hamed:2015bza}%
  \BibitemOpen
  \bibfield  {author} {\bibinfo {author} {\bibfnamefont {N.}~\bibnamefont
  {Arkani-Hamed}}\ and\ \bibinfo {author} {\bibfnamefont {J.}~\bibnamefont
  {Maldacena}},\ }\bibfield  {title} {\bibinfo {title} {{Cosmological Collider
  Physics}},\ }\href@noop {} {\  (\bibinfo {year} {2015})},\ \Eprint
  {https://arxiv.org/abs/1503.08043} {arXiv:1503.08043 [hep-th]} \BibitemShut
  {NoStop}%
\bibitem [{\citenamefont {Arkani-Hamed}\ \emph {et~al.}(2020)\citenamefont
  {Arkani-Hamed}, \citenamefont {Baumann}, \citenamefont {Lee},\ and\
  \citenamefont {Pimentel}}]{Arkani-Hamed:2018kmz}%
  \BibitemOpen
  \bibfield  {author} {\bibinfo {author} {\bibfnamefont {N.}~\bibnamefont
  {Arkani-Hamed}}, \bibinfo {author} {\bibfnamefont {D.}~\bibnamefont
  {Baumann}}, \bibinfo {author} {\bibfnamefont {H.}~\bibnamefont {Lee}},\ and\
  \bibinfo {author} {\bibfnamefont {G.}~\bibnamefont {Pimentel}},\ }\bibfield
  {title} {\bibinfo {title} {{The Cosmological Bootstrap: Inflationary
  Correlators from Symmetries and Singularities}},\ }\href
  {https://doi.org/10.1007/JHEP04(2020)105} {\bibfield  {journal} {\bibinfo
  {journal} {JHEP}\ }\textbf {\bibinfo {volume} {04}},\ \bibinfo {pages}
  {105}},\ \Eprint {https://arxiv.org/abs/1811.00024} {arXiv:1811.00024
  [hep-th]} \BibitemShut {NoStop}%
\bibitem [{\citenamefont {Baumann}\ \emph {et~al.}(2020)\citenamefont
  {Baumann}, \citenamefont {Duaso~Pueyo}, \citenamefont {Joyce}, \citenamefont
  {Lee},\ and\ \citenamefont {Pimentel}}]{Baumann:2019oyu}%
  \BibitemOpen
  \bibfield  {author} {\bibinfo {author} {\bibfnamefont {D.}~\bibnamefont
  {Baumann}}, \bibinfo {author} {\bibfnamefont {C.}~\bibnamefont
  {Duaso~Pueyo}}, \bibinfo {author} {\bibfnamefont {A.}~\bibnamefont {Joyce}},
  \bibinfo {author} {\bibfnamefont {H.}~\bibnamefont {Lee}},\ and\ \bibinfo
  {author} {\bibfnamefont {G.}~\bibnamefont {Pimentel}},\ }\bibfield  {title}
  {\bibinfo {title} {{The Cosmological Bootstrap: Weight-Shifting Operators and
  Scalar Seeds}},\ }\href {https://doi.org/10.1007/JHEP12(2020)204} {\bibfield
  {journal} {\bibinfo  {journal} {JHEP}\ }\textbf {\bibinfo {volume} {12}},\
  \bibinfo {pages} {204}},\ \Eprint {https://arxiv.org/abs/1910.14051}
  {arXiv:1910.14051 [hep-th]} \BibitemShut {NoStop}%
\bibitem [{\citenamefont {Baumann}\ \emph {et~al.}(2021)\citenamefont
  {Baumann}, \citenamefont {Duaso~Pueyo}, \citenamefont {Joyce}, \citenamefont
  {Lee},\ and\ \citenamefont {Pimentel}}]{Baumann:2020dch}%
  \BibitemOpen
  \bibfield  {author} {\bibinfo {author} {\bibfnamefont {D.}~\bibnamefont
  {Baumann}}, \bibinfo {author} {\bibfnamefont {C.}~\bibnamefont
  {Duaso~Pueyo}}, \bibinfo {author} {\bibfnamefont {A.}~\bibnamefont {Joyce}},
  \bibinfo {author} {\bibfnamefont {H.}~\bibnamefont {Lee}},\ and\ \bibinfo
  {author} {\bibfnamefont {G.}~\bibnamefont {Pimentel}},\ }\bibfield  {title}
  {\bibinfo {title} {{The Cosmological Bootstrap: Spinning Correlators from
  Symmetries and Factorization}},\ }\href
  {https://doi.org/10.21468/SciPostPhys.11.3.071} {\bibfield  {journal}
  {\bibinfo  {journal} {SciPost Phys.}\ }\textbf {\bibinfo {volume} {11}},\
  \bibinfo {pages} {071} (\bibinfo {year} {2021})},\ \Eprint
  {https://arxiv.org/abs/2005.04234} {arXiv:2005.04234 [hep-th]} \BibitemShut
  {NoStop}%
\bibitem [{\citenamefont {Sleight}(2020)}]{Sleight:2019mgd}%
  \BibitemOpen
  \bibfield  {author} {\bibinfo {author} {\bibfnamefont {C.}~\bibnamefont
  {Sleight}},\ }\bibfield  {title} {\bibinfo {title} {{A Mellin Space Approach
  to Cosmological Correlators}},\ }\href
  {https://doi.org/10.1007/JHEP01(2020)090} {\bibfield  {journal} {\bibinfo
  {journal} {JHEP}\ }\textbf {\bibinfo {volume} {01}},\ \bibinfo {pages}
  {090}},\ \Eprint {https://arxiv.org/abs/1906.12302} {arXiv:1906.12302
  [hep-th]} \BibitemShut {NoStop}%
\bibitem [{\citenamefont {Sleight}\ and\ \citenamefont
  {Taronna}(2020)}]{Sleight:2019hfp}%
  \BibitemOpen
  \bibfield  {author} {\bibinfo {author} {\bibfnamefont {C.}~\bibnamefont
  {Sleight}}\ and\ \bibinfo {author} {\bibfnamefont {M.}~\bibnamefont
  {Taronna}},\ }\bibfield  {title} {\bibinfo {title} {{Bootstrapping
  Inflationary Correlators in Mellin Space}},\ }\href
  {https://doi.org/10.1007/JHEP02(2020)098} {\bibfield  {journal} {\bibinfo
  {journal} {JHEP}\ }\textbf {\bibinfo {volume} {02}},\ \bibinfo {pages}
  {098}},\ \Eprint {https://arxiv.org/abs/1907.01143} {arXiv:1907.01143
  [hep-th]} \BibitemShut {NoStop}%
\bibitem [{\citenamefont {Pajer}(2021)}]{Pajer:2020wxk}%
  \BibitemOpen
  \bibfield  {author} {\bibinfo {author} {\bibfnamefont {E.}~\bibnamefont
  {Pajer}},\ }\bibfield  {title} {\bibinfo {title} {{Building a Boostless
  Bootstrap for the Bispectrum}},\ }\href
  {https://doi.org/10.1088/1475-7516/2021/01/023} {\bibfield  {journal}
  {\bibinfo  {journal} {JCAP}\ }\textbf {\bibinfo {volume} {01}},\ \bibinfo
  {pages} {023}},\ \Eprint {https://arxiv.org/abs/2010.12818} {arXiv:2010.12818
  [hep-th]} \BibitemShut {NoStop}%
\bibitem [{\citenamefont {Goodhew}\ \emph {et~al.}(2021)\citenamefont
  {Goodhew}, \citenamefont {Jazayeri},\ and\ \citenamefont
  {Pajer}}]{Goodhew:2020hob}%
  \BibitemOpen
  \bibfield  {author} {\bibinfo {author} {\bibfnamefont {H.}~\bibnamefont
  {Goodhew}}, \bibinfo {author} {\bibfnamefont {S.}~\bibnamefont {Jazayeri}},\
  and\ \bibinfo {author} {\bibfnamefont {E.}~\bibnamefont {Pajer}},\ }\bibfield
   {title} {\bibinfo {title} {{The Cosmological Optical Theorem}},\ }\href
  {https://doi.org/10.1088/1475-7516/2021/04/021} {\bibfield  {journal}
  {\bibinfo  {journal} {JCAP}\ }\textbf {\bibinfo {volume} {04}},\ \bibinfo
  {pages} {021}},\ \Eprint {https://arxiv.org/abs/2009.02898} {arXiv:2009.02898
  [hep-th]} \BibitemShut {NoStop}%
\bibitem [{\citenamefont {Jazayeri}\ \emph {et~al.}(2021)\citenamefont
  {Jazayeri}, \citenamefont {Pajer},\ and\ \citenamefont
  {Stefanyszyn}}]{Jazayeri:2021fvk}%
  \BibitemOpen
  \bibfield  {author} {\bibinfo {author} {\bibfnamefont {S.}~\bibnamefont
  {Jazayeri}}, \bibinfo {author} {\bibfnamefont {E.}~\bibnamefont {Pajer}},\
  and\ \bibinfo {author} {\bibfnamefont {D.}~\bibnamefont {Stefanyszyn}},\
  }\bibfield  {title} {\bibinfo {title} {{From Locality and Unitarity to
  Cosmological Correlators}},\ }\href {https://doi.org/10.1007/JHEP10(2021)065}
  {\bibfield  {journal} {\bibinfo  {journal} {JHEP}\ }\textbf {\bibinfo
  {volume} {10}},\ \bibinfo {pages} {065}},\ \Eprint
  {https://arxiv.org/abs/2103.08649} {arXiv:2103.08649 [hep-th]} \BibitemShut
  {NoStop}%
\bibitem [{\citenamefont {Di~Pietro}\ \emph {et~al.}(2022)\citenamefont
  {Di~Pietro}, \citenamefont {Gorbenko},\ and\ \citenamefont
  {Komatsu}}]{DiPietro:2021sjt}%
  \BibitemOpen
  \bibfield  {author} {\bibinfo {author} {\bibfnamefont {L.}~\bibnamefont
  {Di~Pietro}}, \bibinfo {author} {\bibfnamefont {V.}~\bibnamefont
  {Gorbenko}},\ and\ \bibinfo {author} {\bibfnamefont {S.}~\bibnamefont
  {Komatsu}},\ }\bibfield  {title} {\bibinfo {title} {{Analyticity and
  Unitarity for Cosmological Correlators}},\ }\href
  {https://doi.org/10.1007/JHEP03(2022)023} {\bibfield  {journal} {\bibinfo
  {journal} {JHEP}\ }\textbf {\bibinfo {volume} {03}},\ \bibinfo {pages}
  {023}},\ \Eprint {https://arxiv.org/abs/2108.01695} {arXiv:2108.01695
  [hep-th]} \BibitemShut {NoStop}%
\bibitem [{\citenamefont {Hogervorst}\ \emph {et~al.}(2023)\citenamefont
  {Hogervorst}, \citenamefont {Penedones},\ and\ \citenamefont
  {Vaziri}}]{Hogervorst:2021uvp}%
  \BibitemOpen
  \bibfield  {author} {\bibinfo {author} {\bibfnamefont {M.}~\bibnamefont
  {Hogervorst}}, \bibinfo {author} {\bibfnamefont {J.}~\bibnamefont
  {Penedones}},\ and\ \bibinfo {author} {\bibfnamefont {K.~S.}\ \bibnamefont
  {Vaziri}},\ }\bibfield  {title} {\bibinfo {title} {{Towards the
  Non-Perturbative Cosmological Bootstrap}},\ }\href
  {https://doi.org/10.1007/JHEP02(2023)162} {\bibfield  {journal} {\bibinfo
  {journal} {JHEP}\ }\textbf {\bibinfo {volume} {02}},\ \bibinfo {pages}
  {162}},\ \Eprint {https://arxiv.org/abs/2107.13871} {arXiv:2107.13871
  [hep-th]} \BibitemShut {NoStop}%
\bibitem [{\citenamefont {Pimentel}\ and\ \citenamefont
  {Wang}(2022)}]{Pimentel:2022fsc}%
  \BibitemOpen
  \bibfield  {author} {\bibinfo {author} {\bibfnamefont {G.}~\bibnamefont
  {Pimentel}}\ and\ \bibinfo {author} {\bibfnamefont {D.-G.}\ \bibnamefont
  {Wang}},\ }\bibfield  {title} {\bibinfo {title} {{Boostless Cosmological
  Collider Bootstrap}},\ }\href {https://doi.org/10.1007/JHEP10(2022)177}
  {\bibfield  {journal} {\bibinfo  {journal} {JHEP}\ }\textbf {\bibinfo
  {volume} {10}},\ \bibinfo {pages} {177}},\ \Eprint
  {https://arxiv.org/abs/2205.00013} {arXiv:2205.00013 [hep-th]} \BibitemShut
  {NoStop}%
\bibitem [{\citenamefont {Jazayeri}\ and\ \citenamefont
  {Renaux-Petel}(2022)}]{Jazayeri:2022kjy}%
  \BibitemOpen
  \bibfield  {author} {\bibinfo {author} {\bibfnamefont {S.}~\bibnamefont
  {Jazayeri}}\ and\ \bibinfo {author} {\bibfnamefont {S.}~\bibnamefont
  {Renaux-Petel}},\ }\bibfield  {title} {\bibinfo {title} {{Cosmological
  bootstrap in slow motion}},\ }\href {https://doi.org/10.1007/JHEP12(2022)137}
  {\bibfield  {journal} {\bibinfo  {journal} {JHEP}\ }\textbf {\bibinfo
  {volume} {12}},\ \bibinfo {pages} {137}},\ \Eprint
  {https://arxiv.org/abs/2205.10340} {arXiv:2205.10340 [hep-th]} \BibitemShut
  {NoStop}%
\bibitem [{\citenamefont {Wang}\ \emph {et~al.}(2023)\citenamefont {Wang},
  \citenamefont {Pimentel},\ and\ \citenamefont {Ach\'ucarro}}]{Wang:2022eop}%
  \BibitemOpen
  \bibfield  {author} {\bibinfo {author} {\bibfnamefont {D.-G.}\ \bibnamefont
  {Wang}}, \bibinfo {author} {\bibfnamefont {G.}~\bibnamefont {Pimentel}},\
  and\ \bibinfo {author} {\bibfnamefont {A.}~\bibnamefont {Ach\'ucarro}},\
  }\bibfield  {title} {\bibinfo {title} {{Bootstrapping Multi-Field Inflation:
  Non-Gaussianities from Light Scalars Revisited}},\ }\href
  {https://doi.org/10.1088/1475-7516/2023/05/043} {\bibfield  {journal}
  {\bibinfo  {journal} {JCAP}\ }\textbf {\bibinfo {volume} {05}},\ \bibinfo
  {pages} {043}},\ \Eprint {https://arxiv.org/abs/2212.14035} {arXiv:2212.14035
  [astro-ph.CO]} \BibitemShut {NoStop}%
\bibitem [{\citenamefont {Arkani-Hamed}\ \emph {et~al.}(2017)\citenamefont
  {Arkani-Hamed}, \citenamefont {Benincasa},\ and\ \citenamefont
  {Postnikov}}]{Arkani-Hamed:2017fdk}%
  \BibitemOpen
  \bibfield  {author} {\bibinfo {author} {\bibfnamefont {N.}~\bibnamefont
  {Arkani-Hamed}}, \bibinfo {author} {\bibfnamefont {P.}~\bibnamefont
  {Benincasa}},\ and\ \bibinfo {author} {\bibfnamefont {A.}~\bibnamefont
  {Postnikov}},\ }\bibfield  {title} {\bibinfo {title} {{Cosmological Polytopes
  and the Wavefunction of the Universe}},\ }\href@noop {} {\  (\bibinfo {year}
  {2017})},\ \Eprint {https://arxiv.org/abs/1709.02813} {arXiv:1709.02813
  [hep-th]} \BibitemShut {NoStop}%
\bibitem [{\citenamefont {Benincasa}(2018)}]{Benincasa:2018ssx}%
  \BibitemOpen
  \bibfield  {author} {\bibinfo {author} {\bibfnamefont {P.}~\bibnamefont
  {Benincasa}},\ }\bibfield  {title} {\bibinfo {title} {{From the Flat-Space
  S-matrix to the Wavefunction of the Universe}},\ }\href@noop {} {\  (\bibinfo
  {year} {2018})},\ \Eprint {https://arxiv.org/abs/1811.02515}
  {arXiv:1811.02515 [hep-th]} \BibitemShut {NoStop}%
\bibitem [{\citenamefont {Benincasa}(2019)}]{Benincasa:2019vqr}%
  \BibitemOpen
  \bibfield  {author} {\bibinfo {author} {\bibfnamefont {P.}~\bibnamefont
  {Benincasa}},\ }\bibfield  {title} {\bibinfo {title} {{Cosmological Polytopes
  and the Wavefunction of the Universe for Light States}},\ }\href@noop {} {\
  (\bibinfo {year} {2019})},\ \Eprint {https://arxiv.org/abs/1909.02517}
  {arXiv:1909.02517 [hep-th]} \BibitemShut {NoStop}%
\bibitem [{\citenamefont {Raju}(2012)}]{Raju:2012zr}%
  \BibitemOpen
  \bibfield  {author} {\bibinfo {author} {\bibfnamefont {S.}~\bibnamefont
  {Raju}},\ }\bibfield  {title} {\bibinfo {title} {{New Recursion Relations and
  a Flat Space Limit for AdS/CFT Correlators}},\ }\href
  {https://doi.org/10.1103/PhysRevD.85.126009} {\bibfield  {journal} {\bibinfo
  {journal} {Phys. Rev. D}\ }\textbf {\bibinfo {volume} {85}},\ \bibinfo
  {pages} {126009} (\bibinfo {year} {2012})},\ \Eprint
  {https://arxiv.org/abs/1201.6449} {arXiv:1201.6449 [hep-th]} \BibitemShut
  {NoStop}%
\bibitem [{\citenamefont {Arkani-Hamed}\ \emph {et~al.}()\citenamefont
  {Arkani-Hamed}, \citenamefont {Baumann}, \citenamefont {Hillman},
  \citenamefont {Joyce}, \citenamefont {Lee},\ and\ \citenamefont
  {Pimentel}}]{Companion}%
  \BibitemOpen
  \bibfield  {author} {\bibinfo {author} {\bibfnamefont {N.}~\bibnamefont
  {Arkani-Hamed}}, \bibinfo {author} {\bibfnamefont {D.}~\bibnamefont
  {Baumann}}, \bibinfo {author} {\bibfnamefont {A.}~\bibnamefont {Hillman}},
  \bibinfo {author} {\bibfnamefont {A.}~\bibnamefont {Joyce}}, \bibinfo
  {author} {\bibfnamefont {H.}~\bibnamefont {Lee}},\ and\ \bibinfo {author}
  {\bibfnamefont {G.}~\bibnamefont {Pimentel}},\ }\href@noop {} {\emph
  {\bibinfo {title} {{\rm Differential Equations for Cosmological Correlators
  (to appear)}}}}\BibitemShut {NoStop}%
\bibitem [{\citenamefont {Parke}\ and\ \citenamefont
  {Taylor}(1986)}]{Parke:1986gb}%
  \BibitemOpen
  \bibfield  {author} {\bibinfo {author} {\bibfnamefont {S.}~\bibnamefont
  {Parke}}\ and\ \bibinfo {author} {\bibfnamefont {T.}~\bibnamefont {Taylor}},\
  }\bibfield  {title} {\bibinfo {title} {{An Amplitude for $n$ Gluon
  Scattering}},\ }\href {https://doi.org/10.1103/PhysRevLett.56.2459}
  {\bibfield  {journal} {\bibinfo  {journal} {Phys. Rev. Lett.}\ }\textbf
  {\bibinfo {volume} {56}},\ \bibinfo {pages} {2459} (\bibinfo {year}
  {1986})}\BibitemShut {NoStop}%
\bibitem [{\citenamefont {Benincasa}\ and\ \citenamefont
  {Cachazo}(2007)}]{Benincasa:2007xk}%
  \BibitemOpen
  \bibfield  {author} {\bibinfo {author} {\bibfnamefont {P.}~\bibnamefont
  {Benincasa}}\ and\ \bibinfo {author} {\bibfnamefont {F.}~\bibnamefont
  {Cachazo}},\ }\bibfield  {title} {\bibinfo {title} {{Consistency Conditions
  on the S-Matrix of Massless Particles}},\ }\href@noop {} {\  (\bibinfo {year}
  {2007})},\ \Eprint {https://arxiv.org/abs/0705.4305} {arXiv:0705.4305
  [hep-th]} \BibitemShut {NoStop}%
\bibitem [{\citenamefont {Hodges}(2012)}]{Hodges:2012ym}%
  \BibitemOpen
  \bibfield  {author} {\bibinfo {author} {\bibfnamefont {A.}~\bibnamefont
  {Hodges}},\ }\bibfield  {title} {\bibinfo {title} {{A Simple Formula for
  Gravitational MHV Amplitudes}},\ }\href@noop {} {\  (\bibinfo {year}
  {2012})},\ \Eprint {https://arxiv.org/abs/1204.1930} {arXiv:1204.1930
  [hep-th]} \BibitemShut {NoStop}%
\bibitem [{\citenamefont {Cheung}\ \emph {et~al.}(2015)\citenamefont {Cheung},
  \citenamefont {Kampf}, \citenamefont {Novotny},\ and\ \citenamefont
  {Trnka}}]{Cheung:2014dqa}%
  \BibitemOpen
  \bibfield  {author} {\bibinfo {author} {\bibfnamefont {C.}~\bibnamefont
  {Cheung}}, \bibinfo {author} {\bibfnamefont {K.}~\bibnamefont {Kampf}},
  \bibinfo {author} {\bibfnamefont {J.}~\bibnamefont {Novotny}},\ and\ \bibinfo
  {author} {\bibfnamefont {J.}~\bibnamefont {Trnka}},\ }\bibfield  {title}
  {\bibinfo {title} {{Effective Field Theories from Soft Limits of Scattering
  Amplitudes}},\ }\href {https://doi.org/10.1103/PhysRevLett.114.221602}
  {\bibfield  {journal} {\bibinfo  {journal} {Phys. Rev. Lett.}\ }\textbf
  {\bibinfo {volume} {114}},\ \bibinfo {pages} {221602} (\bibinfo {year}
  {2015})},\ \Eprint {https://arxiv.org/abs/1412.4095} {arXiv:1412.4095
  [hep-th]} \BibitemShut {NoStop}%
\bibitem [{\citenamefont {Arkani-Hamed}\ \emph {et~al.}(2018)\citenamefont
  {Arkani-Hamed}, \citenamefont {Bai}, \citenamefont {He},\ and\ \citenamefont
  {Yan}}]{Arkani-Hamed:2017mur}%
  \BibitemOpen
  \bibfield  {author} {\bibinfo {author} {\bibfnamefont {N.}~\bibnamefont
  {Arkani-Hamed}}, \bibinfo {author} {\bibfnamefont {Y.}~\bibnamefont {Bai}},
  \bibinfo {author} {\bibfnamefont {S.}~\bibnamefont {He}},\ and\ \bibinfo
  {author} {\bibfnamefont {G.}~\bibnamefont {Yan}},\ }\bibfield  {title}
  {\bibinfo {title} {{Scattering Forms and the Positive Geometry of Kinematics,
  Color and the Worldsheet}},\ }\href {https://doi.org/10.1007/JHEP05(2018)096}
  {\bibfield  {journal} {\bibinfo  {journal} {JHEP}\ }\textbf {\bibinfo
  {volume} {05}},\ \bibinfo {pages} {096}},\ \Eprint
  {https://arxiv.org/abs/1711.09102} {arXiv:1711.09102 [hep-th]} \BibitemShut
  {NoStop}%
\bibitem [{\citenamefont {Cachazo}\ \emph {et~al.}(2014)\citenamefont
  {Cachazo}, \citenamefont {He},\ and\ \citenamefont {Yuan}}]{Cachazo:2013iea}%
  \BibitemOpen
  \bibfield  {author} {\bibinfo {author} {\bibfnamefont {F.}~\bibnamefont
  {Cachazo}}, \bibinfo {author} {\bibfnamefont {S.}~\bibnamefont {He}},\ and\
  \bibinfo {author} {\bibfnamefont {E.}~\bibnamefont {Yuan}},\ }\bibfield
  {title} {\bibinfo {title} {{Scattering of Massless Particles: Scalars, Gluons
  and Gravitons}},\ }\href {https://doi.org/10.1007/JHEP07(2014)033} {\bibfield
   {journal} {\bibinfo  {journal} {JHEP}\ }\textbf {\bibinfo {volume} {07}},\
  \bibinfo {pages} {033}},\ \Eprint {https://arxiv.org/abs/1309.0885}
  {arXiv:1309.0885 [hep-th]} \BibitemShut {NoStop}%
\bibitem [{\citenamefont {Arkani-Hamed}\ \emph {et~al.}(2023)\citenamefont
  {Arkani-Hamed}, \citenamefont {Frost}, \citenamefont {Salvatori},
  \citenamefont {Plamondon},\ and\ \citenamefont
  {Thomas}}]{Arkani-Hamed:2023lbd}%
  \BibitemOpen
  \bibfield  {author} {\bibinfo {author} {\bibfnamefont {N.}~\bibnamefont
  {Arkani-Hamed}}, \bibinfo {author} {\bibfnamefont {H.}~\bibnamefont {Frost}},
  \bibinfo {author} {\bibfnamefont {G.}~\bibnamefont {Salvatori}}, \bibinfo
  {author} {\bibfnamefont {P.-G.}\ \bibnamefont {Plamondon}},\ and\ \bibinfo
  {author} {\bibfnamefont {H.}~\bibnamefont {Thomas}},\ }\bibfield  {title}
  {\bibinfo {title} {{All Loop Scattering as a Counting Problem}},\ }\href@noop
  {} {\  (\bibinfo {year} {2023})},\ \Eprint {https://arxiv.org/abs/2309.15913}
  {arXiv:2309.15913 [hep-th]} \BibitemShut {NoStop}%
\end{thebibliography}%

\end{document}